\documentclass[a4paper,11pt]{article}
\usepackage{jcappub}
\usepackage[T1]{fontenc}
\usepackage{subcaption}
\usepackage{graphicx}
\usepackage{amsmath}
\usepackage{amssymb}
\usepackage{lipsum}
\usepackage{comment}
\usepackage{multirow}
\usepackage{xcolor}
\usepackage{hyperref}
\usepackage[normalem]{ulem}
\usepackage{mathtools}
\usepackage{bm}
\usepackage{bbold}
\usepackage{soul}
\usepackage{import}
\pdfoutput=1

\usepackage{aas_macros}

\newcommand{\beq}{\begin{equation}}
\newcommand{\beqa}{\begin{eqnarray}}
\newcommand{\eeq}{\end{equation}}
\newcommand{\eeqa}{\end{eqnarray}}

\newcommand{\vect}[1]{\mbox{\boldmath${#1}$}}

\newcommand{\lla}{\left\langle}

\newcommand{\rra}{\right\rangle}

\newcommand{\vex}{{\vect x}}

\newcommand{\ven}{\vect n}

\newcommand{\ve}{{\vect e}}
\def\be{\begin{equation}}
\def\ee{\end{equation}}
\def\bi{\begin{itemize}}
\def\ei{\end{itemize}}
\def\ben{\begin{enumerate}}
\def\een{\end{enumerate}}

\title{\centering \fontsize{19}{11}\selectfont Boltzmann equations for astrophysical\\Stochastic Gravitational Wave Backgrounds\\scattering off of massive objects}

\author[a,d]{Lorenzo Pizzuti}
\author[b]{, Alessandro Tomella}
\author[c]{, Carmelita Carbone}
\author[d]{, Matteo Calabrese}
\author[e,f,g]{, Carlo Baccigalupi}

\affiliation[a]{CEICO, Institute of Physics of the Czech Academy of Sciences, Na Slovance 2, 182 21 Praha 8, Czechia}
\affiliation[b]{ Dipartimento di Fisica “Aldo Pontremoli”, Università degli Studi di Milano, Via Celoria 16, I-20133 Milano, Italy}
\affiliation[c]{ INAF -- Istituto di Astrofisica Spaziale e Fisica cosmica di Milano (IASF-MI), Via Alfonso Corti 12, I-20133 Milano, Italy}
\affiliation[d]{Astronomical Observatory of the Autonomous Region of the Aosta Valley (OAVdA), Loc.
Lignan 39, I-11020, Nus (Aosta Valley), Italy}
\affiliation[e]{SISSA, Via Bonomea 265, 34136 Trieste,  Italy}
\affiliation[f]{ IFPU -- Institute for fundamental physics of the  Universe, Via Beirut 2, 34014 Trieste, Italy}
\affiliation[g]{ INFN -- Sezione di Trieste, via Valerio 2, 34127 Trieste, Italy}

\emailAdd{pizzuti@fzu.cz}
\emailAdd{alessandro.tomella@studenti.unimi.it}
\emailAdd{carmelita.carbone@inaf.it}
\emailAdd{calabrese@oavda.it}
\emailAdd{carlo.baccigalupi@sissa.it}

\abstract{
The goal of  this work is to present a set of coupled Boltzmann equations describing the intensity and polarisation Stokes parameters of the SGWB.  {Collision terms (as discussed e.g. in Ref.~\cite{Cusin}) which account} for gravitational Compton scattering off of massive objects, are also included. This set resembles that for the CMB Stokes parameters, but the different spin nature of the gravitational radiation and the physics involved in the scattering process determine crucial differences. In the case of gravitational Compton scattering, due to the Rutherford angular dependence of the cross section, all the SGWB intensity multipoles of order $\ell$ are scattered out, therefore producing outgoing intensity anisotropies of any order $\ell$ if they are present in the incoming radiation. On the other hand, {as already outlined in~\cite{Cusin}, SGWB linear polarisation modes can be expanded in a basis of spherical harmonics with $m=\pm 4$ and $\ell\ge 4$}. This means that SGWB polarisation modes can be generated from unpolarised anisotropic radiation only with $m=\pm 4$, therefore requiring at least a hexadecapole anisotropy ($\ell\ge 4$) in the incoming intensity. {Assuming a simplified toy model where scattering targets are localised in a small redshift range, we solve analytically the set of coupled Boltzmann equations to get explicit expressions for the intensity and polarisation angular power spectra}. We confirm the contribution of the gravitational Compton scattering to the SGWB anisoptropies is extremely small for collisions with massive compact objects (BH and SMBH) in the frequency range of current and upcoming surveys. The system of coupled Boltzmann equations presented here provides a way to accurate estimate the total amount of anisotropies generated by multiple SGWB scattering processes off of massive objects, as well as the interplay between polarisation and intensity, during the GW propagation across the LSS of the universe. These results will be useful for the full treatment of the astrophysical SWGB anisotropies in view of upcoming gravitational waves observatories.}

\begin{document}
\maketitle
\flushbottom

\section{Introduction}
The recent direct observations of Gravitational Wave (GW) signals~\cite{Abbott2016, Abbott2017a, Abbott2017b, Abbott2021a, Abbott2021b, LKV2021, Abbott2021d, Abbott2021e, Hernandez2019, NANOGrav2020} have opened a new exciting era for astronomical and cosmological studies. The increase in sensitivity of current and upcoming detectors \cite{Maggiore2020,Baker2019,Seoane2022,Evans2021,Kawamura2019}, both ground-based (e.g. the Laser Interferometer Gravitational Observatory, LIGO\footnote{\url{https://www.ligo.caltech.edu/}}, the VIRGO interferometer\footnote{\url{https://www.virgo-gw.eu/}},  the Kamioka Gravitational Wave Detector, KAGRA\footnote{\url{https://gwcenter.icrr.u-tokyo.ac.jp/en/}}, the Einstein Telescope, ET\footnote{\url{http://www.et-gw.eu}}, the Cosmic Explorer, CE\footnote{\url{https://cosmicexplorer.org/}}), and space-based (e.g. the  Laser Interferometer Space Antenna, LISA\footnote{\url{https://www.lisamission.org/}}, and the Deci-hertz Interferometer Gravitational wave Observatory, DECIGO\footnote{\url{https://decigo.jp/index_E.html}}) will provide measurements of the Stochastic Gravitational Wave Background (SGWB) of both astrophysical and cosmological origins. In this context, a complete characterisation of the astrophysical SGWB and its evolution is crucial to distinguish it from the cosmological SGWB and obtain fundamental information about the very early  universe. In particular, the analysis of the anisotropies in intensity (e.g.~\cite{Bartolo2019}) and polarisation (e.g.~\cite{Domcke:2019zls, Orlando2022}) of the latter would reveal the details of different physical processes in the inflationary era. 
The study of the SGWB \textit{intensity} \cite{Cusin2017,Cusin2018,Cusin2018b,Jenkins2019,Jenkins2018b,Pitrou2020,Bertacca2020,Jenkins2019c,Jenkins2019d,Allen1997,Jenkins2018,Cusin2022,Capurri1,Capurri2,Boco,Galloni2022} has been performed so far with approaches similar to those used for the Cosmic Microwave Background (CMB): its evolution is described by a phase-space distribution function obeying the collisionless Boltzmann equation in a background perturbed by the Large Scale Structure (LSS) of the  universe~\cite{Contaldi, Bartolo2020,Valbusadallarmi2020,Valbusadallarmi2022,Dimastrogiovanni22}. This approximation applies in the so-called geometrical optics limit \cite{Isaacson1, Isaacson2,Anile&Breuer1974,Gravitation,Maggiore_Vol1}, i.e. when the GW wavelengths are much smaller than the typical spatial scale on which the perturbed background varies. 
As such, on cosmological scales, higher frequency gravitons behave as effectively spin-2 massless particles freely moving along geodesics. However, regarding the collisionless aspect of this treatment, as pointed out by Ref.~\cite{Cusin}, in the proximity of compact or extended massive objects, such as stars or black holes, graviton wavelengths could be much larger than the scale of variation of the spacetime generated by such structures, and wave effects, producing SWGB polarisation, need to be accounted for. In this respect, Ref.~\cite{Cusin} has included the emissivity term in the Boltzmann equation of the SGWB intensity, but neglected the collision term, under the reasonable assumption that scattering effects, and the resultant SGWB polarisation, can be computed discarding the back reaction of polarisation on intensity.

In this work we present a set of Boltzmann equations for both the SGWB \textit{intensity} and \textit{polarisation}, with collision terms able to couple the system equations by including the SGWB scattering off of massive structures. To this aim, we exploit an approach similar to the treatment of CMB photons~\cite{Durrer2008, Hu, Zaldarriaga, Kosowsky95, Kamionkowski}, where both the incoming and the scattered photons are considered free fields and the interaction term is a function of them~\cite{Kosowsky95}. In this case we consider the so-called Born approximation, i.e. we assume the GW propagation and scattering processes as characterised by two different length scales, the mean free path and the scattering length, with the assumption that, for weak enough couplings, the former is much larger than the latter.
Under these hypotheses, for the gravitational scattering from massive objects, in the SGWB Boltzmann equations we consider as collision source the so-called Compton graviton-scalar scattering in the low-energy limit~\cite{Peters, Gross, Guadagnini, Holstein, Dolan}. 

This paper is organised as follows: in Sec.~\ref{sec:Stokes} we report the definitions of the density matrix and Stokes parameters for GW backgrounds; in Sec.~\ref{sec:compton} {we review the main features of the Compton cross-section entering the RHS of the collisional Boltzmann equations, in the case of the scattering of gravitons off of massive objects. We highlight the results obtained in the literature (e.g.~\cite{DeLogi1977, Cusin, Guadagnini, Holstein, Dolan}) and used in our computations; in Sec.~\ref{sec:Boltzmann} we derive the full set of Boltzmann equations for the Stokes parameters of GW backgrounds; in Sec.~\ref{sec:scatter_anis} we analyse the main features of the scattered SGWB intensity and polarisation anisotropies; in Sec.~\ref{sec:solution} we present and discuss the solution for the intensity and polarisation angular power spectra in the case of a simplified, toy-model scenario; finally, in the last Section we draw our conclusions.}  

\section{Density matrix and Stokes parameters for gravitational radiation}
\label{sec:Stokes}
The \textit{gravitational} Stokes parameters (e.g.~\cite{Anile&Breuer1974,Seto06,Gubitosi2018,Conneely18}) provide, in analogy to the electromagnetic case, a way to completely characterise the intensity and polarisation content of the SGWB. The standard plane wave expansion of a generic GW, $h_{ij}$, in the so-called transverse traceless gauge, is given by
\begin{align}
h_{ij}(t,\vex)=\sum_{P=+,\times} \int^{\infty}_{-\infty} \text{d}f \int_{S^2} \text{d}\Omega\,
h_P(f,\ven) e^{2\pi i f (t-\ven \cdot \vex) } \ve^P_{ij}(\ven),
\end{align}
where $S^2$ is the unit sphere for the angular integral, the unit vector $\ven=(\sin\theta\cos\phi, \sin\theta\sin\phi,\cos\theta)$ is the propagation direction,
$\ve^+_{ij}=[{\hat \ve}_\theta \otimes {\hat \ve}_\theta- {\hat \ve}_\phi \otimes  {\hat \ve}_\phi]_{ij}$ and 
$\ve^\times_{ij}=[{\hat \ve}_\theta \otimes {\hat \ve}_\phi+{\hat \ve}_\phi \otimes {\hat \ve}_\theta]_{ij}$
is the basis for transverse-traceless tensors such that $e^A_{ij} {e^B}^{ij} = 2 \delta^{AB}$ ($A,B=+,\times$) where ${\hat \ve}_\theta$ and ${\hat \ve}_\phi$ are two unit vectors with a fixed spherical coordinate system.
As the observed $h_{ij}(t,\vex)$ is real, one has a relation for the complex conjugate: $h_P(-f,\ven)=h_P(f,\ven)^*$. When replacing the direction $\ven\to -\ven$, one has $ \ve^+_{ij} \to \ve^+_{ij}$ (even parity) and $\ve^\times_{ij}\to  -\ve^\times_{ij}$ (odd parity).
The covariance matrix, $\rho (f,\ven)$,
between two polarisation modes, $h_+(f,\ven)$ and $h_\times(f^\prime,\ven^\prime)$, is given in terms of the GW Stokes parameters, $I$, $Q$, $U$, $V$, as~\cite{Seto06}:
\beq
\label{density_matrix}
\rho(f, \ven)=\delta(f,f^\prime)\left( 
           \begin{array}{@{\,}cc@{\,}}
           I+Q & U-iV  \\
            U+iV & I-Q  \\ 
           \end{array} \right) \frac{\delta_{D}(\ven-\ven^\prime)}{4\pi}\,,
\eeq
where
\begin{align}
\label{Stokes_params}
&I(f,\ven)= \lla |h_+(f,\ven)|^2+|h_\times(f,\ven)|^2\rra /2\,, \\ 
&Q(f,\ven)=\lla |h_+(f,\ven)|^2-|h_\times(f,\ven)|^2\rra /2\,, \\ 
&U(f,\ven)= \lla h_+(f,\ven) h_\times(f,\ven)^*+ h_+(f,\ven)^* h_\times(f,\ven) \rra /2\,, \\ 
&V(f,\ven)=i \lla h_+(f,\ven) h_\times(f,\ven)^*- h_+(f,\ven)^* h_\times(f,\ven) \rra /2\,.
\end{align}
From now on, we omit the $f$-dependence for simplicity.
The symbol $\lla \dots\rra$ represents the ensemble average
for a superposition of stationary incoherent waves, and
$\delta_{D}(\cdot)$ is the delta  function on the unit sphere. 
The parameter $I(\ven)$ represents the total intensity of the SGWB, $Q(\ven)$ and $U(\ven)$ are related to its linear polarisation, and $V(\ven)$ to the circular polarisation. 
When the basis vectors, ${\hat \ve}_\theta $ and $ {\hat \ve}_\phi$, are rotated around the propagation direction, $\ven$, by an angle $\gamma$, the parameters $I$
and $V$ are invariant (spin 0), while the combinations $(Q\pm iU)(\ven)$ transform as $(Q\pm iU)^\prime(\ven^\prime)=e^{\mp 4i \gamma}(Q\pm iU)(\ven)$, i.e. they are objects of spin $s=\pm 4$ \cite{radipro}. 

Eq.~\eqref{density_matrix} represents the so-called density matrix, in the linear polarisation
basis ${e_{ij}}^{+/\times}$, of an ensemble of gravitons in mixed state, and can be written in short notation as:
\begin{equation}
\rho=\left(I\mathbb{1}+Q\sigma_3+U\sigma_1+V\sigma_2\right)/2,
\end{equation}
where $\mathbb{1}$ is the identity matrix, and $\sigma_i$ are the Pauli spin matrices. 
Thus the density matrix for an ensemble of gravitons contains the same information as the four Stokes parameters, and the time evolution of the density matrix provides the time evolution of the system.

In the considered quasi-classical limit, and for astrophysical SGWB, one can assume that the graviton phase-space distribution function, $f(x^\mu,p^\mu)$ (treated in Sec.~\ref{sec:Boltzmann} below), is given by the Wigner distribution function, i.e. the Wigner transform of the density matrix~\cite{Wigner1932, Wigner84, Wigner1984b, Calzetta1988, Calzetta1988b, Calzetta1989, Gong&Seo2020}. Under such a hypothesis, in the following we will assume that, as for CMB photons, perturbations to the graviton density matrix correspond to perturbations to the classical gravitational Stokes parameters describing the intensity and polarisation of the astrophysical SGWB, and behaving as phase-space distribution functions which obey the respective Boltzmann equation with a collision term given by the gravitational Compton scattering from massive objects in the low energy limit~\cite{Guadagnini}.

\section{Compton graviton-scalar scattering in the low energy limit}
\label{sec:compton}
The process of GW scattering off of compact and massive objects, treated as spin-0 quantities, has been studied since the sixties via several methods, both from quantum-mechanical~\cite{Gross, Holstein, Guadagnini} and classical~\cite{Westervelt71, Peters, Dolan, DeLogi1977} perspectives. Here {we review results in the literature (e.g.~\cite{DeLogi1977, Cusin, Guadagnini, Holstein, Dolan}), which in turn will enter the collision term in the RHS of the SGWB Boltzmann equations representing the real novelty of this work}. In particular, we will consider the \textit{long wavelength/weak-field} limit of gravitational Compton scattering, i.e. when the wavelength, $\lambda$, of the incoming wave is assumed to be much larger than the Schwarzschild radius, $r_s=2GM$, of the scattering object of mass $M$, i.e. $\pi r_s/\lambda =M \omega \ll 1$~\cite{Dolan}. We use units of $c=1$ and, in first approximation, neglect any possible angular momentum of the compact object since it has been suggested
that, to first order in $M \omega$, gravitational waves do not exhibit angular-momentum-induced effects~\cite{DeLogi1977}. 

In particular, considering a gravitational wave with polarisation tensor $\hat \epsilon_{ij}^\prime$ hitting, from the direction $\textbf n^\prime$, a compact object of mass $M$ and zero angular momentum, the differential cross section, in the TT gauge and in the rest frame of the object, is \cite{DeLogi1977}
\begin{equation} \label{eq:cross-sec2}
    \frac{d\sigma_{\epsilon ^\prime \epsilon}}{d\Omega} = \frac{(GM)^2}{\sin^4 \beta/2}|\hat \epsilon_{ij}^\prime \hat \epsilon^{ij}|^2 \,,
\end{equation}
where ${\hat \epsilon}^{ij}$ is the outgoing polarisation of the wave, and $\beta$ is the angle between the incoming and outgoing directions.

In the so-called \textit{scattering plane geometry}, one can decompose the polarisation basis in terms of a pair of unit vectors normal to the direction of propagation, $\hat l$ for the vector in the scattering plane and $\hat m$ for the vector normal to it, and write
\begin{align}
     & \hat{\epsilon}_{ij}^+ = \hat l_i\hat l_j - \hat m_i\hat m_j \nonumber \\
    & \hat{\epsilon}_{ij}^\times = \hat l_i \hat m_j + \hat m_i \hat l_j\,.
\end{align}
Using a similar decomposition for the incoming polarisation tensor $\hat \epsilon^\prime_{ij}$, and the relation between the ``primed'' and ``unprimed'' vectors in terms of the incoming direction $\textbf{n}^\prime$:
\begin{align}\label{eq: ScatteringAngle}
    & \textbf{l} \cdot \textbf{l}^\prime = \textbf{n} \cdot \textbf{n}^\prime= \cos\beta \,,\nonumber \\
    & \textbf{m} \cdot \textbf{m}^\prime = 1,
\end{align}
one then obtains the relations
\begin{align}\label{eq:ScatteredPolarization}
    & \hat{\epsilon}_{ij}^+ \, \hat{\epsilon}^{+^\prime ij} = 1 + \cos^2\beta \nonumber \\
    & \hat{\epsilon}_{ij}^\times \, \hat{\epsilon}^{\times ^\prime ij}  = 2\cos \beta\, \nonumber \\
    & \hat{\epsilon}_{ij}^+ \, \hat{\epsilon}^{\times ^\prime ij} = \hat{\epsilon}_{ij}^\times \, \hat{\epsilon}^{+^\prime ij} = 0 \,.
\end{align}
Finally, using Eqs.~\eqref{eq:cross-sec2}-\eqref{eq:ScatteredPolarization}, one can derive the polarisation-summed cross section averaged over the initial polarisation states, $e^\prime_{ij}({\bf n}^\prime)$ (unpolarised incident GWs), and summed over the final states, $e_{ij}({\bf n})$~\cite{DeLogi1977, Cusin, Guadagnini, Holstein, Dolan}
\begin{align}\label{tot_angural_dependence}
   & \frac{d\sigma}{d\Omega} = \frac{1}{2}\frac{(GM)^2}{\sin^4 \beta/2}\sum_{A = +, \times}\sum_{B = +, \times} |\hat \epsilon_{ij}^A \hat \epsilon^{B^\prime ij}|^2 = \nonumber \\
   & =   \frac{(GM)^2}{8}\frac{1 + 6\cos^2\beta+\cos^4\beta}{\sin^4(\beta/2)} \,
 \end{align}
However, here we are interested in investigating how the GW components $h_+$ and $h_\times$ get modified in the scattering event. It comes out that, similarly to the electromagnetic case, the plus and cross amplitudes need to be multiplied by their respective factors in Eq.~\eqref{eq:ScatteredPolarization}, but a polarisation-independent angular dependence is also present, such that
\begin{align}\label{eq:ScatteredAmplitudes}
   & h_+ = \frac{GM}{\sin^2\beta/2} h_{ij}^\prime \hat{\epsilon}_{ij}^+ = GM \frac{(1+\cos^2\beta)}{\sin^2\beta/2} \,h_+^\prime \nonumber \\
   & h_\times = \frac{GM}{\sin^2\beta/2} h_{ij}^\prime \hat{\epsilon}_{ij}^\times = GM \frac{2\cos\beta}{\sin^2\beta/2} \,h_\times^\prime \,.
\end{align}

Note that, unlike the electromagnetic case, in gravitational wave scattering the helicity is not conserved and the scattering cross section is nonzero in the backward direction, implying the existence of a helicity-reversing amplitude \cite{DeLogi1977, Dolan}. Moreover, Eq.~\eqref{tot_angural_dependence} diverges as $\beta^{-4}$ for small scattering angles, i.e. in the forward scattering direction. This feature is the gravitational equivalent of the Rutherford's angular dependence in electromagnetism, and, as pointed out in Refs.~\cite{Peters, Bjerrum-Bohr2015, Cusin, Gross}, it is related to the long range nature of the gravitational interaction, or, alternatively, to the non-linear nature of Einstein's equations~\cite{Holstein}. 

In this work, in order to fix this divergence, we follow {and summarise in this Section the method proposed by Ref.~\cite{Cusin}} which introduces a natural cutoff scale on $\beta$, in order to deal with a well-defined \textit{total cross-section} of the process. {In particular, as pointed out in Ref.~\cite{Cusin},} this is a standard procedure when considering Boltzmann equations for systems where long range interactions are involved, such as the Coulomb scattering in plasma (e.g. Ref.~\cite{He21} and references therein).
The cutoff scale comes from the observation that wave-like effects are non-negligible only within a region whose size is given by the square root of the Kretschmann scalar, $K$, associated to the metric generated by the scattering object. Assuming this is well approximated by a Schwarzschild metric, one has $\sqrt{K}=2\sqrt{3}r_\text{s}/r^3$, where $r$ is the distance from the centre of the object. Therefore, for wavelengths $\lambda^{-2} \lesssim 2\sqrt{3}r_\text{s}/r^3$, i.e. $r \lesssim (2\sqrt{3}r_\text{s}\lambda^2)^{1/3} \equiv R_\lambda$, wave-effects are non-negligible. The region, defined by $R_\lambda$, around the scattering object, depends on both its size and the wavelength of the incoming GW. In the far-field regime, the relation between the scattering angle, $\beta$, and the impact parameter, $b$, of a graviton geodesic, is given by $\beta \approx 2r_s/b$. Therefore, if an upper bound exists for $b$, one can find a corresponding lower bound on $\beta$. The idea proposed by Ref.~\cite{Cusin} is to choose $b_{\text{max}} = R_\lambda$, so that the minimum value of the scattering angle is given by $ \beta_\text{min} = 2\left(r_s/R_\lambda\right) = \left[4 r_s^2/(\sqrt 3 \lambda^2)\right]^{1/3}\,$.
Analogously, Ref.~\cite{Cusin} provides also an upper bound $\beta_\text{max}\approx 2$ for the impact parameter of compact objects, which  translates in a lower bound $b_\text{min}$. Moreover, assuming that for sources corresponding to black holes, wave effects occur in a region $r_s < r < R_\lambda$, implying $r_s<R_\lambda$, it follows that
a limit for the observed and redshifted wavelength, $\lambda_\text{obs}=(1+z)\lambda$, of a GW scattered by a massive object of mass $M$ at redshift $z$, is given by
\begin{equation}\label{eq:lambdalim}
    \lambda_\text{obs}\ge 10^{-13}\left(\frac{M}{M_\odot}\right)(1+z) \,\text{pc}.
\end{equation}
Thus, GWs with relatively small wavelengths are affected only by scattering due to small mass objects, while GWs with large $\lambda$ receive  contribution from massive objects as well. In Tab.~\ref{tab:lambdas} we summarise the scattering masses of compact sources which fall in the observational ranges of ongoing and upcoming GW experiments.
\begin{table} 
\centering
\resizebox{0.8\textwidth}{!}{%
\begin{tabular}{|l|r|r|r|}
\hline
\fontsize{9}{11}\selectfont 
Detector & $f_\text{obs}\, [\text{Hz}]$ & $\lambda_\text{obs}\, [\text{pc}]$ & Mass  $[M_\odot]$  \\
\hline
\fontsize{8}{11}\selectfont
LIGO/VIRGO & $10^1 - 10^4$ & $10^{-12} - 10^{-9}$ & $\lesssim 1 - 10^3$ \\ 
\fontsize{8}{11}\selectfont
ET & $10^0 - 10^4$ & $10^{-12} - 10^{-8}$ & $\lesssim 1 - 10^4$ \\ 
\fontsize{8}{11}\selectfont
LGWA & $10^{-3} - 10^0$ & $10^{-8} - 10^{-5}$ & $\lesssim 10^4 - 10^7$ \\ 
\fontsize{8}{11}\selectfont
LISA & $10^{-5} - 10^{0}$ &  $10^{-8} - 10^{-3}$ &  $\lesssim 10^4 - 10^9$ \\ 
\fontsize{8}{11}\selectfont
PTA & $10^{-9} - 10^{-6}$ &  $10^{-2} - 10^{1}$ &  $\lesssim 10^{10}$ \\ 
\hline
\end{tabular}
}
\caption[range of masses]{\label{tab:lambdas} Values of the mass of compact objects whose scattering contribution affects GWs in the frequency ranges of ongoing and upcoming detectors: advanced LIGO/VIRGO (e.g.~\cite{aLigo2015,Acernese_2014}), Einstein Telescope (ET~\cite{Punturo_2010, Hild2011}), LISA~\cite{LISA17}, Lunar Gravitational Wave Antenna (LGWA~\cite{LGWA21})  and Pulsar Timing Arrays (PTA, e.g.~\cite{PTA}). We assume the scattering to occur at an average redshift $z = 1$. Note that, for $f\lesssim 10^{-4}\,\text{Hz}$ all objects in a reasonable range of masses (star-like objects, intermediate mass BH, supermassive BH) contribute to the GW scattering.}
\end{table}

Finally, adopting the above limits on the scattering angle and integrating Eq.~\eqref{tot_angural_dependence} over the solid angle, one obtains the total cross-section:
\begin{equation} \label{eq:totalcross}
    \sigma(\lambda, M) = (GM)^2 \mathcal T(\lambda, M) ,
\end{equation}
where
\begin{equation}\label{integrated_section}
    \mathcal T(\lambda, M)\equiv \frac{1}{8}\int_0^{2\pi}\text{d}\varphi \int_{\beta_{\text{min}(\lambda)}}^{\beta_{\text{max}}} \text{d}\beta \sin\beta \frac{1 + 6\cos^2\beta+\cos^4\beta}{\sin^4\beta/2} \,.
\end{equation}

\section{SGWB Boltzmann equations with collision terms}\label{sec:Boltzmann}

In the quasi-classical limit, the general approach to describe the evolution of the SGWB, travelling across the LSS of the  universe, passes through the graviton Wigner function, i.e. the graviton phase-space distribution function, $f(x^\mu,p^\mu)$, where $x^\mu$ denotes the spacetime coordinates, $p^\mu=\text{d}x^\mu/d\lambda$ is the four-momentum, and $\lambda$ is an affine parameter along the graviton trajectories. This distribution function follows the Boltzmann equation (e.g.~\cite{Contaldi,Bartolo2020,Valbusadallarmi2020}):
\begin{equation}\label{eq:boltz1}
    \mathcal{L}[f]=\mathcal{C}[f]+\mathcal{J}[f],
\end{equation}
where $\mathcal{L}=\text{d}/\text{d}\lambda$ is the Liouville operator, $\mathcal{C}[f]$ is the collision term representing the interaction between GW and massive objects, and $\mathcal{J}[f]$ is a source term due to emission processes from astrophysical and cosmological GW sources. Here, we are interested in evaluating,  in the Born approximation, $\mathcal{C}[f]$ which contributes both to the SGWB intensity and polarisation, while we ignore the contribution from the emissivity term, $\mathcal{J}[f]$~\cite{Bartolo2020, Cusin}, and possible effects of graviton scattering with other particles (see e.g.~\cite{Bartolo18, Valbusadallarmi2020}). 

The approach followed in this work is based on the analogy with the standard treatment of CMB anisotropies~\cite{Durrer2008, Hu, Kamionkowski}. In particular, we use the fact that, in the photon propagation description, the vector radiative transfer equation (VRTE) - which describes the rate of change of the Stokes parameters - is formally equivalent to the Boltzmann equations for CMB intensity and polarisation (see e.g.~\cite{Kosowsky95}). We apply the same treatment to the GW Stokes parameters defined via the density matrix in Eq.~\eqref{density_matrix}, for which, as the system becomes classical, $f(x^\mu,p^\mu)$ can be interpreted as its Wigner transform.

We therefore consider the GWs \textit{Stokes vector} $\vec{\bm {\mathcal{S}}}=(I,\,Q+iU,\,Q-iU,\,V)^\text{T}$ and write the Boltzmann equation as~\cite{Zaldarriaga,Hu, Durrer2008}
\begin{equation}\label{eq:VRTE}
    \frac{d\vec{\bm {\mathcal{S}}}}{d\eta}= \dot{\tau}(\eta) \Big[\vec{\bm{\mathcal F}}\big[\vec{\bm{\mathcal S}}\big]-\vec{\bm{\mathcal{S}}}\Big]\,,
\end{equation}
where $\eta$ is the conformal time, $\vec{\bm{\mathcal F}}\big[\vec{\bm{\mathcal S}}\big]$ is the flux of the outgoing radiation, and $\tau(\eta)$ defines the "optical" thickness, in this case due to graviton scattering, of the medium through which the gravitational radiation is propagating. The dot indicates derivative with respect to $\eta$. In the case of gravitational waves incoming over a distribution of massive objects, the optical thickness is given by~\cite{Cusin}:
\begin{equation}
\label{eq:optical_depth}
    \tau(\eta)=\int^{\eta_0}_\eta n_\text{ph}(\eta^\prime)\sigma(\eta^\prime,\lambda,M)a(\eta^\prime)\text{d}\eta^\prime \,,
\end{equation}
for an initial conformal time $\eta_0$. In the above equation, $n_\text{ph}(\eta)$ is the physical density of massive objects which incoming GWs are scattered off at a given epoch $\eta$, $a(\eta)$ is the scale factor of the  universe, and $\sigma$ is the total cross-section of Eq.~\eqref{eq:totalcross}.
Note that here we work in a simplified scenario of incoherent scattering~\cite{Cusin} assuming all the massive scatters to have the same mass $M$. In principle, one could refine this approximation by integrating, over reasonable mass ranges, the R.H.S of Eq.~\eqref{eq:VRTE} times a weighting function $f(M)$ that describes the mass distribution of the objects. However, as we will see in Sec.~\ref{sec:scatter_anis}, this will not affect the overall results of our analysis.  

After scattering, the outgoing radiation is given by:
\begin{equation}\label{scattering_contr}
    \vec{\bm{\mathcal F}}\big[\vec{\bm{\mathcal S}}\big] = \frac{1}{\mathcal T}\int \text{d}\Omega^\prime \mathcal M(\textbf n^\prime, \textbf n )\vec{\bm{\mathcal S}}(\textbf n^\prime),
\end{equation}
with $\textbf n^\prime$ and $\textbf n$ the incident and outgoing directions, respectively.
All the physics of the scattering process is described by the \textit{scattering matrix}, $\mathcal M$, which connects the incoming and the outgoing Stokes vectors. We derive an explicit expression for it in terms of \textit{Spin Weighted Spherical Harmonics} (SWSH), $_sY_ {\ell m}({\bf n})$, with spin $s=0,\pm 4$ and order $\ell\le 4$ (see Appendix ~\ref{app:spinw} and e.g.~\cite{Goldberg67,Durrer2008}):
\begin{equation} \label{eq:GravitonScatteringMatrix}
\begin{split}
&\mathcal M(\textbf n, \textbf n^\prime) = \frac{4}{(1 - \textbf n \cdot \textbf n^\prime )^2}\left[\frac{8}{5}\mathcal{P}^{(0)} + \frac{64\pi}{35}\sum_{m = -2}^2\mathcal P^{(2)}_m(\textbf n, \textbf n^\prime) + \right.\\
& \left. +\frac{16\pi}{315}\sum_{m = -4}^4\mathcal P^{(4)}_m(\textbf n, \textbf n^\prime)+\frac{16\pi}{105}\left(\sum_{m=-1}^1\mathcal P_m^{(1)}(\textbf n, \textbf n^\prime)+\sum_{m=-3}^3\mathcal P_m^{(3)}(\textbf n, \textbf n^\prime)\right)\right]\,,
\end{split}
\end{equation}
where we have included circular polarisation terms ($l=1,3$) and made explicit the dependence on scalar and spin weighted spherical harmonics:
\begin{equation}
\begin{split}
&\mathcal P^{(0)} = \text{diag}(1, 0, 0, 0) \,, \\
&\mathcal P_m^{(1)}(\textbf n, \textbf n^\prime)=\text{diag}\Big( 0,0,0,\,70\,Y_{1m}(\textbf n) Y^*_{1m}( \textbf n^\prime) \Big)\,,\\
&\mathcal P^{(2)}_m(\textbf n, \textbf n^\prime) =  \text{diag}(\,Y_{2m}(\textbf n) Y^*_{2m}( \textbf n^\prime)\, , 0, 0,0)\,,\\
& \mathcal P_m^{(3)}(\textbf n, \textbf n^\prime)=\text{diag}\left(0,0,0,\,3\sqrt{7}\,Y_{3m}(\textbf n) Y^*_{3m}( \textbf n^\prime)\right)\,,\\
&\mathcal P^{(4)}_m(\textbf n, \textbf n^\prime) = 
\begin{pmatrix}
Y_{4m}(\textbf n)Y^*_{4m}(\textbf n^\prime) & \sqrt\frac{35}{2}\,Y_{4m}(\textbf n){}_4Y^*_{4m}(\textbf n^\prime) & \sqrt\frac{35}{2} \,Y_{4m}(\textbf n){}_{-4}Y^*_{4m}(\textbf n^\prime) & 0\\
    \sqrt {70} \, {}_4Y_{4m}(\textbf n)Y^*_{4m}(\textbf n^\prime) & 35\,\,{}_4Y_{4m}(\textbf n){}_4Y^*_{4m}(\textbf n^\prime) & 35\,\,{}_4Y_{4m}(\textbf n){}_{-4}Y^*_{4m}(\textbf n^\prime) & 0 \\
    \sqrt{70}\,{}_{-4}Y_{4m}(\textbf n)Y^*_{4m}(\textbf n^\prime) & 35\,\,{}_{-4}Y_{4m}(\textbf n){}_4 Y^*_{4m}(\textbf n^\prime) & 35\,\,{}_{-4}Y_{4m}(\textbf n){}_{-4}Y^*_{4m}(\textbf n^\prime) & 0 \\
    0 & 0 & 0 & 0
\end{pmatrix}\,.
\end{split}
\end{equation}
Computation details are given in Appendix \ref{app:matrix}. A quick inspection of the null terms in the structure of $\mathcal M$ reveals that the circular polarisation is not generated by the gravitational Compton scattering for an incident SGWB background without an initial circular polarisation; this is a feature in common with Thomson scattering between electrons and CMB photons~\cite{DeLogi1977, Dolan, Kosowsky95,Cusin}.
However, we highlight here three crucial differences with the CMB Thomson scattering process: 
\begin{itemize}

    \item First, the gravitational Compton scattering matrix is characterised by the presence of the overall factor $(1-\textbf{n}\cdot \textbf{n}^\prime)^{-2}$ coming from the Rutherford's type denominator of the cross-section, Eq.~\eqref{eq:cross-sec2}, {which complicates the angular dependence as compared to the CMB case.}
    
    \item Second,  the presence of SWSH of order 4 in $\mathcal{M}({\bf n}, {\bf n^\prime})$ {is expected since $Q \pm iU$ transform under rotations as spin-4 quantities, thus they have to be decomposed in terms of spin-4 weighted spherical harmonics:}
    \begin{equation}
        (Q\pm i U)(\textbf n) =\sum_{\ell,m}a_{\ell m}^{\pm 4}{}_{\pm 4}Y_{\ell,m}(\textbf n)\,.
    \end{equation}
    Instead, in the photon case, the linear polarisations are spin-2 quantities (see e.g.~\cite{Zaldarriaga}), thus the CMB scattering matrix contains SWSH of $s\pm 2$ \cite{Durrer2008}. This is related to the nature of the propagating field (spin-1 in the case of photons and spin-2 in the case of gravitons). 

    \item Third, in $\mathcal P^{(4)}_m(\textbf n, \textbf n^\prime)$ we see terms of order $\ell = 4$ (hexadecapole) in scalar and spin-weighted spherical harmonics (instead of the quadrupole, $\ell = 2$, that characterises
CMB Thomson scattering); this is a consequence of the spin-0 nature of the intensity, $I$, and, again, the spin-4 nature of the polarisation, $Q\pm iU$, in the graviton case.
\end{itemize}

\subsection{The perturbed Boltzmann equations for gravitational Stokes parameters}
\label{subsec:Boltzmann_equations}
We can now insert the scattering matrix, $\mathcal M(\textbf n, \textbf n^\prime)$ of Eq.~\eqref{eq:GravitonScatteringMatrix}, into the Boltzmann equation~\eqref{eq:VRTE}, and expand it to first order in terms of graviton energy and polarisation vectors\footnote{The approach of dealing with density matrices and distributions functions is completely different from the usual one of considering GWs as linear metric perturbations and solve for their evolution by linearising the Einstein's equations. Therefore, the graviton ensemble described by the density matrix perturbed at first order in energy and polarisation vectors can represent GWs beyond the linear regime in terms of the standard theory of cosmological perturbations.} to obtain the evolution equation of the gravitational Stokes parameters in an inhomogeneous  universe. For the sake of simplicity, in the LHS (Liouville part) of the Boltzmann equations~\eqref{eq:VRTE} we consider only linear scalar metric perturbations over a flat FLRW background and neglect possible tensor metric perturbations, in order to avoid double counting (unless separated in distinct frequency ranges such that the geometrical optics limit still holds) the tensor perturbations in the SGWB described by the phase-space distribution function, $f(x^\mu,p^\mu)$, travelling on top of the perturbed background. In the conformal Newtonian gauge, the line element reads:  
\begin{equation}\label{eq:metric}
\text{d}s^2=a^2(\eta)[(1+2\Psi)\text{d}\eta^2-(1-2\Phi)\delta_{ij}\text{d}x^i\text{d}x^j]\,,
\end{equation}
where $\Psi$ and $\Phi$ are $\ll 1$ and functions of the spacetime. The metric perturbations generate fluctuations, $\rho_{ij}^{(1)}$, to the unperturbed graviton density matrix, $\rho_{ij}^{(0)}$: 
\begin{equation}
\rho_{ij}(\eta, {\bf x},q,{\bf n})\simeq\rho_{ij}^{(0)}(\eta, q)+\rho_{ij}^{(1)}(\eta, {\bf x},q,{\bf n}) \,,
\end{equation}
where, $q = a|{\bf p}|$ is the comoving momentum modulus, $\rho_{ij}^{(0)}$ represents the \textit{uniform} and \textit{unpolarised} GW density matrix such that $\rho_{11}^{(0)}=\rho_{22}^{(0)}, \rho_{12}^{(0)}=\rho_{21}^{(0)}=0$, and $\rho_{ij}^{(1)}=\rho_{ji}^{(1)}$ is the linear perturbation of Eq.~\eqref{density_matrix}.
Therefore we expand at linear order the Stokes vector $\vec{\bm {\mathcal{S}}}$, associated to the density matrix $\rho_{ij}$, and, following the approach of Ref.~\cite{Kosowsky95}, we define the linear perturbations of the SGWB Stokes parameters as  
\begin{equation}\label{rho_pert}
    \begin{split}
        & \Delta_I=\left(\frac{q}{4}\frac{\partial \rho_{11}^{(0)}}{\partial q}\right)^{-1}\left[\rho_{11}^{(1)}+\rho_{22}^{(1)}\right]\,,\\
        & \Delta_{Q\pm iU} =\left(\frac{q}{4}\frac{\partial \rho_{11}^{(0)}}{\partial q}\right)^{-1}\left[\rho_{11}^{(1)}-\rho_{22}^{(1)}\pm i(\rho_{12}^{(1)}+\rho_{21}^{(1)})\right]\,,\\
        & \Delta_{V} =-i\left(\frac{q}{4}\frac{\partial \rho_{11}^{(0)}}{\partial q}\right)^{-1}\left[\rho_{12}^{(1)}-\rho_{21}^{(1)}\right]\,.      
    \end{split}
\end{equation}
In the following, for simplicity of notation we retain only the $\textbf n$ dependence of all the quantities. In Section \ref{sec:solution} we explicitly recall the full dependence when needed.
The above quantities represents linear order fluctuations in the Stokes parameters conveniently scaled by the energy density of the background graviton distribution. To obtain the perturbed R.H.S of the SGWB Boltzmann equation, Eq.~\eqref{eq:VRTE}, we substitute Eq.~\eqref{eq:GravitonScatteringMatrix} into Eq.~\eqref{scattering_contr} and perturb the Stokes vector, $\vec{\bm {\mathcal{S}}}$, at linear order in terms of the quantities defined in Eq.~\eqref{rho_pert}. Moreover, similarly to \cite{Cusin}, we introduce the following integrals, making here the dependence on the SWSH explicit:
\begin{equation}\label{integrals}
\begin{split}
& \mathcal I^{(0)}(\textbf{n}) = \frac{1}{\mathcal T}\frac{32}{5} \int \frac{\text{d}\Omega^\prime}{(1- \textbf n \cdot\textbf n^\prime)^2 }\Delta_I(\textbf n^\prime) \,,\\
& \mathcal I^{(2)}_m(\textbf{n}) = \frac{1}{\mathcal T}\frac{256\pi}{35}\int \frac{\text{d}\Omega^\prime}{(1 - \textbf{n}^\prime\cdot\textbf{n})^2} \, \Delta_I(\textbf{n}^\prime)Y^*_{2m}(\textbf{n}^\prime) \,,\\
& \mathcal I^{(4)}_m(\textbf{n}) = \frac{1}{\mathcal T}\frac{64\pi}{315}\int \frac{\text{d}\Omega^\prime}{(1 - \textbf{n}^\prime\cdot\textbf{n})^2} \, \Delta_I(\textbf{n}^\prime)Y^*_{4m}(\textbf{n}^\prime) \,,\\
& \mathcal L^{(4)\, \pm}_m(\textbf{n}) = \frac{1}{\mathcal T}\frac{64\pi}{315}\int \frac{\text{d}\Omega^\prime}{(1 - \textbf{n}^\prime\cdot\textbf{n})^2} \, \Delta_{Q\pm iU}(\textbf{n}^\prime){}_{\pm4}Y^*_{4m}(\textbf{n}^\prime) \,,\\
& \mathcal V^{(1)}_m(\textbf{n}) =\frac{1}{\mathcal T} \frac{32 \pi}{3}\int \frac{\text{d}\Omega^\prime}{(1 - \textbf{n}^\prime\cdot\textbf{n})^2} \, \Delta_V(\textbf{n}^\prime)Y^*_{1m}(\textbf{n}^\prime) \,,\\
& \mathcal V^{(3)}_m(\textbf{n}) = \frac{1}{\mathcal T}\frac{16 \pi}{5\sqrt 7}\int \frac{\text{d}\Omega^\prime}{(1 - \textbf{n}^\prime\cdot\textbf{n})^2} \, \Delta_V(\textbf{n}^\prime)Y^*_{3m}(\textbf{n}^\prime) \,.
\end{split}
\end{equation}
The above quantities can be interpreted as averages over all directions of the incoming intensity and polarisation, weighted by the Rutherford's type factor, $1/(1-\textbf{n}\cdot\textbf{n}^\prime)^{-2}$, and by SWSH with $s=0,\pm4$ and  $\ell=0,1,2,3,4$.
Note that the integrals are not evaluated on the whole solid angle, but they are limited by the restriction given by the cutoff $\beta_{\rm min}<\beta<\beta_{\rm max}$.
Finally, to obtain the full set of linear Boltzmann equations for GWs, we consider Eq.~\eqref{eq:metric} and expand, up to first order in the scalar metric perturbations, the L.H.S of Eq.~\eqref{eq:VRTE}, obtaining the usual Liouville-like term as for the CMB photon distribution function\footnote{In this treatment we are neglecting first order corrections to the cross-section due to bulk velocities of the massive objects and to their angular momentum (see e.g.~\cite{Guadagnini}).} (e.g. \cite{Kosowsky,Durrer2008, Bartolo2020, Kamionkowski}).

We recall here we are working under two different assumptions: the geometrical optics limit, i.e. GW wavelengths being much smaller than the typical spatial scale on which linear metric perturbations vary, and the Born approximation, i.e. the mean free path of gravitons scattered off by massive objects being much larger than the scattering length.

Finally, the final set of linear Boltzmann equations for the gravitational Stokes parameters of the astrophysical SGWB read as\\ 
\newline
\textbf{Intensity equation:}
\begin{equation}\label{eq:IntensityEquation}
\resizebox{1.\textwidth}{!}{$
\begin{split}
(\partial_\eta  + n^i\partial_i) \Delta_I(\textbf n ) - 4\left[\dot \Phi - n^i\partial_i \Psi\right] =& - a\sigma n_\text{ph}\Big[ \Delta_I(\textbf{n})  - \mathcal I^{(0)}(\textbf{n}) - \sum_{m=-2}^2 Y_{2m}(\textbf n)\mathcal I^{(2)}_m(\textbf{n}) - \sum_{m = -4}^4 Y_{4m}(\textbf n )\times \\
&\Big( \mathcal I^{(4)}_m(\textbf{n}) + \sqrt\frac{35}{2}\mathcal L^{(4)\, +}_m(\textbf{n}) + \sqrt\frac{35}{2}\mathcal L^{(4)\, -}_m(\textbf{n})\Big) \Big]
\end{split}
$}
\end{equation}
\textbf{Linear polarisation equation:}
\begin{equation}\label{eq:PolarizationEquation}
\resizebox{1.\textwidth}{!}{$
(\partial_\eta  + n^i\partial_i) \Delta_{Q\pm iU}(\textbf{n}) =   - a\sigma n_\text{ph}\Big[ \Delta_{Q\pm iU}(\textbf{n}) 
- \sum_{m = -4}^4 {}_{\pm 4} Y_{4m}(\textbf n ) \Big( \sqrt{70} \mathcal I^{(4)}_m(\textbf{n}) 
+ 35 \mathcal L^{(4)\, +}_m(\textbf{n}) + 35 \mathcal L^{(4)\, -}_m(\textbf{n})\Big) \Big]
$}
\end{equation}
\textbf{Circular polarisation equation:}
\begin{equation}\label{eq:CircularPolarizationEquation}
\resizebox{1.\textwidth}{!}{$
(\partial_\eta  + n^i\partial_i) \Delta_V(\textbf n ) =   - a\sigma n_\text{ph}\Big[ \Delta_V(\textbf{n})
- \sum_{m = -1}^1 Y_{1m}(\textbf n ) \mathcal V^{(1)}_m(\textbf{n}) - \sum_{m = -3}^3 Y_{3m}(\textbf n ) \mathcal V^{(3)}_m(\textbf{n})\Big]
$}
\end{equation}
Here we have exploited the fact that the solution of Eq.\eqref{eq:VRTE} at zeroth order in perturbation theory is simply the background uniform unpolarised density matrix, i.e. the redshifted intensity, $\rho_{11}^{(0)}=\rho_{22}^{(0)}$, associated to the monopole of the SGWB. This cancels out between the L.H.S and R.H.S of Eqs.~\eqref{eq:IntensityEquation}-\eqref{eq:CircularPolarizationEquation}.

This system of coupled equations appears rather complicated due to the scattering terms in the R.H.S which include SWSH with $s=\pm 4$ and $\ell\le4$, plus the Rutherford's type factor hidden in the integrals of Eq.~\eqref{integrals}. 
However, a direct inspection of  Eqs.~\eqref{eq:IntensityEquation}-\eqref{eq:CircularPolarizationEquation} reveals some interesting physical features, as described below. 

\section{Intensity and polarisation anistropies of the astrophysical SGWB}
\label{sec:scatter_anis}

In a reference frame where the scattered wave is propagating along the $z$ axis, one can expand the intensity of unpolarised incoming waves in the basis of spherical harmonics:
\begin{equation}
    \tilde{I}'(\beta,\gamma)=\sum_{\ell,m}\tilde{a}_{\ell,m}Y_{\ell,m}(\beta,\gamma)\,,
\end{equation}
where the tilde indicates that quantities are computed in the $z-$aligned frame. The coefficients $\tilde{a}_{\ell,m}$ are related to the $a_{\ell,m}$ in a generic direction as (e.g.~\cite{Kosowsky99}):
\begin{equation}\label{eq:Wignerot}
\tilde{a}_{\ell,m}=\sum_{m^\prime=-\ell}^\ell(\mathcal{D}^*)^\ell_{m^\prime,m}[R(\Theta,\Phi)]a_{\ell,m^\prime}\,.
\end{equation}
In Eq.~\eqref{eq:Wignerot}, the quantities $(\mathcal{D}^*)^\ell_{m^\prime,m}[R(\Theta,\Phi)]$ represent the (complex-conjugate) elements of the Wigner-D matrix (e.g.~\cite{2021PhRvD.103f3528C}), connecting a frame to another one obtained by rotating the former by an angle $\Phi$ along the $z$-axis and an angle $\Theta$ along the $y$-axis.

In the $z$-alinged frame, the scattered intensity $\tilde{I}$ and linear polarisations $\tilde{Q}\pm i \tilde{U}$ are given by (see Eq.~\eqref{eq:matrixRotated} and e.g.~\cite{Cusin}):
\begin{equation}
  \tilde{I}=(GM)^2\int\text{d}\Omega  \tilde{I}^\prime(\beta,\gamma)\frac{1+6\cos^2\beta+\cos^4\beta }{2\sin^4(\beta/2)}\,,
  \label{eq:Iscatter}
\end{equation}
\begin{equation} \label{eq:Qscatter}
    \tilde{Q}\pm i \tilde{U}=(GM)^2\int\text{d}\Omega \tilde{I}^\prime(\beta,\gamma)\frac{\sin^4\beta}{2\sin^4(\beta/2)}e^{\mp4i\gamma}\,.
    \end{equation}
The $\tilde{V}$ Stokes parameter remains everywhere zero if the incoming radiation is initially circularly unpolarised.
Note that, each multipole, $\ell$, with $m\ne 0$ in $I^\prime(\beta,\gamma)$ gives a null contribution to   Eq.~\eqref{eq:Iscatter}, as the integral over $\gamma$ vanishes.

In Thomson scattering for photons, regardless of the anisotropies of the incident radiation, only the monopole $\ell=0$ and the quadrupole $\ell=2$ are affected by the process, while the other multipoles integrate to zero via the orthogonality of spherical harmonics (thus, they remain unchanged with respect to the incident field). 
However, we stress here this is no more true for the gravitational Stokes parameters, since the Rutherford's type divergence, $\sin^{-4}\beta$, causes all the incoming multipoles (with $m=0$) to be scattered, therefore producing outgoing anisotropies of any order $\ell$ if they are present in input.

Regarding $\tilde{Q}\pm i \tilde{U}$, the factor $\exp[\pm i 4\gamma]$ makes the integral in Eq.~\eqref{eq:Qscatter} identically zero unless incident radiation has a non-vanishing $m= \pm 4$ components, which implies $\ell \ge 4$: at least an hexadecapole anisotropy is needed to generate polarisation form an initially unpolarised radiation.

We now estimate the order of magnitude of the R.H.S. of Eqs.~\eqref{eq:IntensityEquation}--\eqref{eq:PolarizationEquation}, by evaluating the cross-section and the expected comoving density of the scattering targets.

Given an observed wavelength, only objects for which the condition Eq.~\eqref{eq:lambdalim} holds contribute to the scattering. For Stellar mass Black Holes (StBH) with typical masses $M\sim 10\,M_\odot$, the wavelengths involved satisfy $\lambda_\text{obs}\gtrsim 10^{-12}\, \text{pc}$, where, in Eq.~\eqref{eq:lambdalim}, we have assumed that the scattering happens at an average redshift $z= 1$. This means that the contribution of StBH ideally affects the observations of all current and upcoming surveys. 
We consider $\lambda_\text{obs}=10^{-10}\,\text{pc}$, which is inside the LIGO/VIRGO band (see Tab.~\ref{tab:lambdas}). From Eq.~\eqref{eq:totalcross}, the total cross-section is 
\begin{equation}
    \sigma(\lambda_\text{obs},10\,M_\odot)=1.47\times 10^{-20}\,\text{pc}^{2}.
\end{equation}

\begin{figure*}
    \centering
    \includegraphics[width=0.9\textwidth]{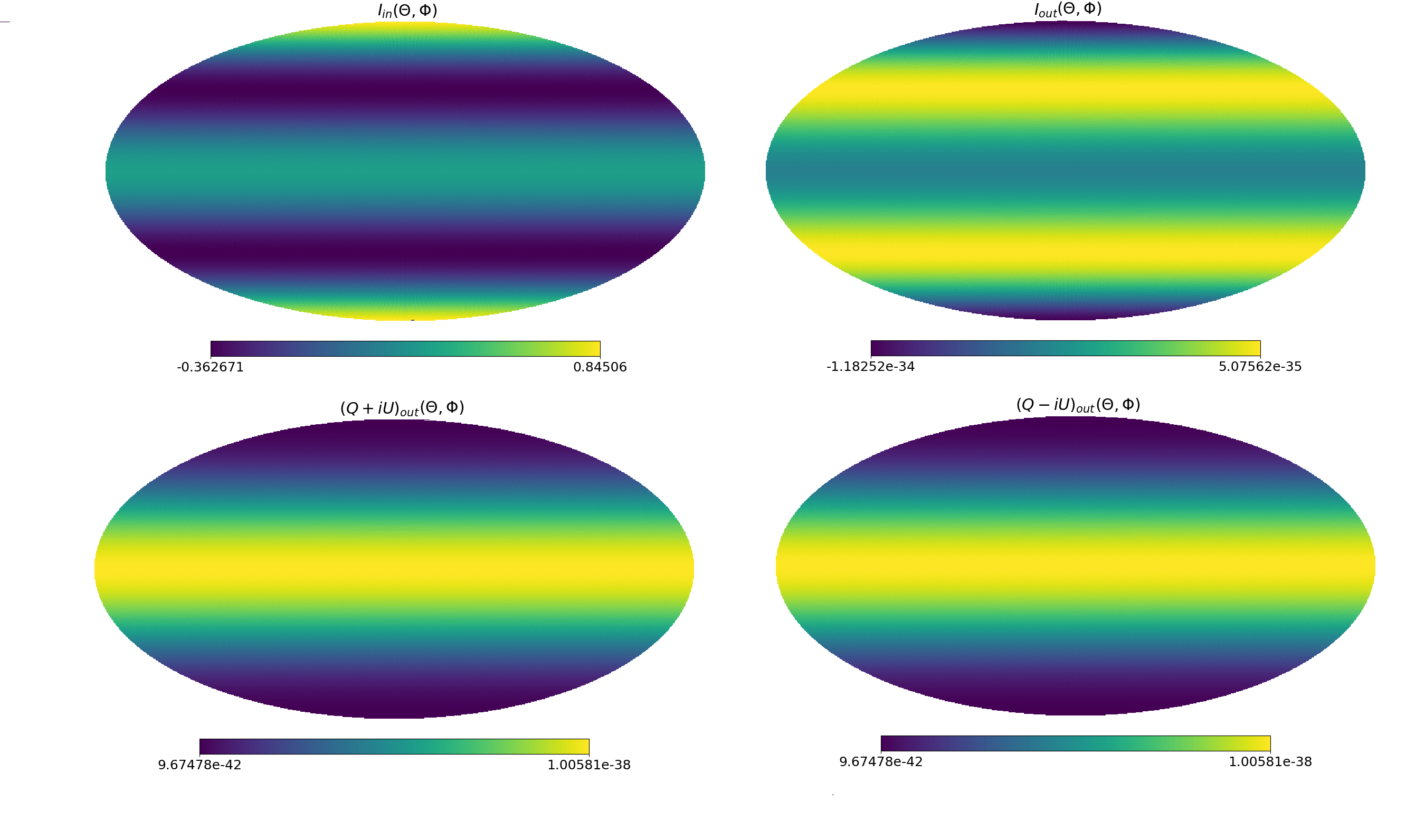}
    \caption{Full-sky maps for the incoming (top left) and outgoing (top right) SGWB intensities, $I$ , and for the outgoing $Q +iU$ (bottom left) and $Q -iU$ (bottom right) polarisations, produced considering the intensity of the incident radiation to be an order-unity pure hexadecapole, $\ell=4,\,m=0$, and assuming absence of incoming polarisation.}
    \label{fig:maps}
\end{figure*}

Following e.g.~\cite{Caputo17,Cusin}, we estimate the number density of StBH in the local  universe to be $n_{\rm ph}=n_{\rm BH}\sim 2.4\times 10^{-14}\,\text{pc}^{-3} $; 
this implies that the overall pre-factor multiplying the R.H.S. of Eqs.~\eqref{eq:IntensityEquation}--\eqref{eq:PolarizationEquation} is extremely small, as one can visually inspect in Fig.~\ref{fig:maps}, where we show the full-sky maps for the outgoing $I_\text{out}$ and $(Q\pm iU)_\text{out}$ Stokes parameters\footnote{We generally refer to  ``outgoing'' radiation as the total contribution in the R.H.S. of Eqs.~\eqref{eq:IntensityEquation}-\eqref{eq:CircularPolarizationEquation}, i.e. with the incoming radiation subtracted.}, as listed in the fourth and fifth columns of Tab.~\ref{tab:ranges}.
The incident radiation (top left panel of Fig.~\ref{fig:maps})  is assumed to be unpolarised and given by a pure intensity hexadecapole, $\ell=4$ with $m=0$, of order-unity:
\begin{equation*}
\begin{split}
   & I_\text{in}(\theta,\phi)=Y_{4,0}(\theta,\phi)=\frac{3}{16\sqrt{\pi}}(35\cos^4\theta-30\cos^2\theta+3),\\
   & (Q\pm i U)_\text{in}(\theta,\phi)=0,
    \end{split}
\end{equation*}
where $(\theta,\phi)$ indicate the angles at which the incoming radiation is observed in the laboratory frame. 

Assuming a value of unity for the incoming intensity, the order-of-magnitude of the collision term is $\sim 10^{-33}$ for the intensity and $\sim 10^{-38}$ for the linear polarisations. 

Thus, as expected, the collision term due to GWs Compton scattering off of massive objects appears to provide a very small contribution to the R.H.S of Boltzmann equations for gravitons, at least when the scattering is sourced by stellar mass BH. 
An order-of-magnitude estimate of the polarisation produced by scattering off of various astrophysical objects can be found in Ref.~\cite{Cusin}, where it is shown that the SGWB polarisation produced by massive scatters is smaller by several orders of magnitude than the intensity anisotropies, for almost all frequency bands of ongoing and upcoming detectors. Nonetheless, we remark that in this work we are not only interested in the total polarisation as in Ref.~\cite{Cusin}, {but rather we focus on the presentation of the system of coupled Boltzmann equations for SGWB intensity and polarisation anisotropies}, which provides a way to estimate the correct, albeit very small, interplay between the two.

Note that an azimuthally-symmetric (i.e. independent of $\phi$) incoming SGWB field with an \textit{hexadecapole} intensity anisotropy  will generate a polarisation in the plane between the incoming and outgoing directions which is real ($U=0$) and proportional to $\sin^4\Theta$, as one can directly verify by using the definitions of the Wigner-D symbols,  and then computing the integrals of Eqs.~\eqref{eq:Iscatter}-\eqref{eq:Qscatter}. This differs from the electromagnetic case, where only a quadrupole intensity anisotropy is needed to generate linear polarisation via Thomson scattering. 

It is important to stress that the integrals of Eq.~\eqref{eq:Iscatter} and Eq.~\eqref{eq:Qscatter} are limited by the cutoff $\beta_\text{min}\propto\lambda_\text{obs}^{-2/3}$. Thus, the larger is the wavelength considered in the scattering process, the smaller is the value of $\beta_\text{min}$. In Tab.~\ref{tab:ranges} we list the values of the optical depth and the corresponding order-of-magnitude values of the collision term for intensity and  polarisation at different $\lambda_\text{obs}$, generated from scattering of order-unity unpolarised incoming radiation. For all the wavelengths, we have considered GW collisions with a population of stellar mass BHs with $M=10 \, M_{\odot}$ at $z\sim 1$. 

Moreover, it is worth noticing that a decrease in $\beta_\text{min}$ does not make the scattered polarisation, Eq.~\eqref{eq:Qscatter}, to diverge, as the integral converges for $\beta\to 0$ (the limit of the ratio $\sin^4\beta/\sin^4(\beta/2)$ in the integrand is finite and equal to $16$ for $\beta\to 0$).
However, Eq.~\eqref{eq:Iscatter} is divergent for small scattering angles, which means that the collisional term in Eq.~\eqref{eq:IntensityEquation} grows with increasing $\lambda$.

{Note that a full numerical solution of the system of coupled SGWB Boltzmann Eqs.~\eqref{eq:IntensityEquation}--\eqref{eq:CircularPolarizationEquation}, would be needed to provide an accurate estimate of the total amount of polarisation generated by multiple scattering processes off of massive objects, as well as its interplay with intensity anisotropies, during the SGWB propagation across the LSS of the universe; this however is beyond the purpose of this paper. 
Nevertheless, it would be interesting to solve the set of equations in a simplified scenario. In the next Section, we provide analytical expressions for the angular power spectrum of intensity and polarisation perturbations, assuming unpolarised incoming radiation scattered off a uniform distribution of targets located in a small redshift interval.}
\begin{table} 
\centering
\resizebox{0.90\columnwidth}{!}{%
\begin{tabular}{|c|c|c|c|c|c|}
\hline
 
$f_\text{obs}\, [\text{Hz}]$ & $\lambda_\text{obs}\, [\text{pc}]$ &  $\dot{\tau}(\lambda,M=10\,M_\odot)$ & R.H.S $(I)$ & R.H.S $(Q\pm i U)$ & Detector \\
\hline
 & &  &  & &\\
$10^2$ & $10^{-10}$ & $ \sim 10^{-33}$ &  $ \sim 10^{-33}$ & $ \sim 10^{-38}$ & LIGO/VIRGO, ET\\ 
$10^{-3}$ & $10^{-5}$ & $ \sim 10^{-27}$ &  $ \sim 10^{-27}$ & $ \sim 10^{-38}$ & LGWA, LISA \\ 
$10^{-7}$ & $10^{-1}$ & $ \sim 10^{-22}$ &  $ \sim 10^{-21}$ & $ \sim 10^{-38}$ & PTA\\ 
$10^{-11}$ & $10^{3}$ & $ \sim 10^{-16}$ &  $ \sim 10^{-16}$ & $ \sim 10^{-38}$ & $--$\\ 
$10^{-14}$ & $10^{6}$ & $ \sim 10^{-12}$ &  $ \sim 10^{-12}$ & $ \sim 10^{-38}$ & $--$\\ 
\hline
\end{tabular}
}
\caption[range of cross-sections]{\label{tab:ranges} Absolute values of the  derivative of the optical depth $\dot{\tau}=-a\sigma n_\text{ph}$ as a function of the observed wavelength. Unpolarised radiation is assumed to scatter off of stellar mass BHs with $M=10\,M_\odot$. The third and fourth columns indicate the order of magnitude of the R.H.S of the Boltzmann equation for the intensity and the polarisation obtained by multiplying Eqs.~\eqref{eq:Iscatter}, \eqref{eq:Qscatter} and the pre-factor $\dot{\tau}/\mathcal{T}$. The last column shows the corresponding detector(s) for which the wavelength falls inside the observational range.}
\end{table}

\section{Intensity and polarisation angular power spectra: a toy model}
\label{sec:solution}
Consider a flux of incoming gravitational radiation propagating since an initial time $\eta_{in}=\eta(z_{in})$ over a cosmological perturbed FLRW background. These gravitational waves are then scattered by a population of massive compact objects within a small interval of redshifts $z\in [1.0,1.05]$, and then freely propagate towards the observer at the present time $\eta_0=\eta(z=0)$.
We can formally solve\footnote{Note that we ignore circular polarisation modes, as they are not sourced in the scattering process.} Eqs.~\eqref{eq:IntensityEquation}-\eqref{eq:PolarizationEquation} by working in Fourier space:
\begin{equation}
\partial_\eta \Delta_I(q,\textbf k,\textbf n,\eta )  + (ik\mu-\dot\tau) \Delta_I(q,\textbf k,\textbf n,\eta) = 4\left[\dot \Phi - ik\mu\Psi\right]-\dot \tau \mathcal{C}_I(q,\textbf k,\textbf n,\eta )\,,
\end{equation}
\begin{equation}
 \partial_\eta \Delta_{Q\pm iU}(q,\textbf k,\textbf n,\eta )  + (ik\mu-\dot\tau) \Delta_{Q\pm iU}(q,\textbf k,\textbf n,\eta) = -\dot \tau \mathcal{C}_{Q\pm iU}(q,\textbf k,\textbf n,\eta )  \,,  
\end{equation}
where we have defined 

\begin{equation*}
  \mathcal{C}_I =  \mathcal I^{(0)} + \sum_{m=-2}^2 Y_{2m}\mathcal I^{(2)}_m + \sum_{m = -4}^4 Y_{4m}\Big( \mathcal I^{(4)}_m + \sqrt\frac{35}{2}\mathcal L^{(4)\, +}_m + \sqrt\frac{35}{2}\mathcal L^{(4)\, -}_m\Big)\,,
\end{equation*}
and 
\begin{equation*}
  \mathcal{C}_{Q\pm iU}=\sum_{m = -4}^4 {}_{\pm 4} Y_{4m} \Big( \sqrt{70} \mathcal I^{(4)}_m 
+ 35 \mathcal L^{(4)\, +}_m + 35 \mathcal L^{(4)\, -}_m\Big)\,.
\end{equation*}
Note that when not needed, we keep as implicit the dependence on the variable ensemble $\{\eta, \textbf k, \textbf n, q\}$. We have further introduced $\mu=\hat{\textbf k} \cdot \hat{\textbf n}$, the cosine between the wavenumber $\textbf k$ and the direction of propagation $\textbf n$.

As in the collisionless case (e.g.~\cite{Bartolo2020} for the case of intenisties anisotropies alone) we can write the formal solution of the Boltzmann equations as:
\begin{equation}
\begin{split}   
\label{eq:solIntensity}
\Delta_I(\eta_0) &=\left[\Delta_I(\eta_{in})+4\Psi(\eta_{in})\right]e^{ik\mu(\eta_{in}-{\eta_0})-\tau(\eta_{in})}+\\
& +\int_{\eta_{in}}^{\eta_{0}}{d\eta e^{ik\mu(\eta-\eta_0)-\tau(\eta)}\left[4(\dot\Phi+\dot\Psi)-\dot\tau(4\Psi+\mathcal{C}_I)\right]}-4\Psi(\eta_0)\,,
\end{split}
\end{equation}
\begin{equation}
\label{eq:solpolarized}
  \Delta_{Q\pm iU}(\eta_0) =\Delta_{Q\pm iU}(\eta_{in})e^{ik\mu(\eta_{in}-{\eta_0})-\tau(\eta_{in})} -\int_{\eta_{in}}^{\eta_{0}}{d\eta e^{ik\mu(\eta-\eta_0)-\tau(\eta)}\dot\tau\mathcal{C}_{Q\pm iU}}\,.
\end{equation}
In both equations we have used the fact that the optical depth in Eq.~\eqref{eq:optical_depth} is zero at $\eta_0$ by definition. The first terms in the RHS carry the information about initial conditions, weighted by the exponential of the optical depth.
We now assume that the initial net polarisation anisotropy of the incoming radiation at $\eta_{in}$ is zero, i.e. the two terms depending on $\mathcal L^{(4)\, \pm}_m$ in $\mathcal{C}_I$ and $\mathcal{C}_{Q\pm iU}$ are zero, as well as the term $\Delta_{Q\pm iU}(\eta_{in})$ in Eq.~\eqref{eq:solpolarized}.
The function $g(\eta)=\dot \tau(\eta)\exp[-\tau(\eta)]$ is called \emph{visibility function} and its integral from $\eta_{in}$ to $\eta_0$ gives the probability that a GW is scattered by massive objects between the initial and present times (see~\cite{Cusin}). We now consider a simplified case in which the scattering targets are well localised in a small redshift interval $[z_{min},z_{max}]$, such that the visibility function is non-vanishing only if $\eta\in[\eta(z_{max}),\eta(z_{min})]$. We can therefore assume that all the other quantities do not change significantly in that time interval and evaluate them at $\eta_*=\eta(z_{max})$.
In this way, since for $\tau(\eta_*)\ll 1$ (see Tab.~\ref{tab:lambdas}) one has:
\begin{equation*}
    \int_{\eta_{in}}^{\eta_{0}}{g(\eta)d\eta } \simeq \int_{\eta_*}^{\eta_{0}}{g(\eta)d\eta} =1-e^{-\tau(\eta_*)}\simeq \frac{\tau(\eta_*)}{1+\tau(\eta_*)}\,,
\end{equation*}
the integrals in the RHS of Eqs.~\eqref{eq:solIntensity}-\eqref{eq:solpolarized} become respectively:
\begin{equation}  
\label{eq:Ineta}
\left[4\Psi(\eta_*)+\mathcal{C}_I(\eta_*)\right]\frac{\tau(\eta_*)}{1+\tau(\eta_*)}e^{ik\mu(\eta_*-\eta_0)}+\int_{\eta_{in}}^{\eta_{0}}{d\eta e^{ik\mu(\eta-\eta_0)-\tau(\eta)}4(\dot\Phi+\dot\Psi)}\,,
\end{equation}
\begin{equation}
\label{eq:Poleta}
  \mathcal{C}_{Q\pm iU}(\eta_*)\frac{\tau(\eta_*)}{1+\tau(\eta_*)}e^{ik\mu(\eta_*-\eta_0)}\,.
\end{equation}

By decomposing the incoming intensity in spherical harmonics
\begin{equation*}
    \Delta_{I}(\eta_*,\textbf k, \textbf n, q)=\sum_{\ell}\sum_{m=-\ell}^\ell \Delta_{I,\ell m}(\eta_*,\textbf k, q)Y_{\ell m}(\textbf n)\,,
\end{equation*}
and using the properties under rotations of Eq.~\eqref{eq:Wignerot} to compute the integrals over the incoming directions in a frame aligned along the $z$-axis, we can rewrite the collision terms $\mathcal{C}_{I}(\eta_*)$, $\mathcal{C}_{Q\pm iU}(\eta_*)$ in Eqs.~\eqref{eq:Ineta}-\eqref{eq:Poleta} in a more useful form:
\begin{equation}
    \mathcal{C}_{I}(\eta_*,\textbf k,\textbf n, q)=\frac{1}{\mathcal T}\sum_{\ell=0}^{\infty}\sum_{m=-\ell}^\ell{\Delta_{I,\ell m}(\eta_*,\textbf k, q)Y_{\ell m}(\textbf n)\mathcal{K}_\ell}\,,
\end{equation}
\begin{equation}
    \mathcal{C}_{Q \pm i U}(\eta_*,\textbf k,\textbf n, q)=16\sqrt{\frac{2}{35}}\frac{1}{\mathcal T}\sum_{\ell=0}^{\infty}\sum_{m=-\ell}^\ell{\Delta_{I,\ell m}(\eta_*,\textbf k, q){}_{\pm 4}Y_{\ell m}(\textbf n)\mathcal{K}_{\ell}^{\mp 4}}\,,
\end{equation}
where we have defined
\begin{equation*}
    \mathcal{K}_{\ell}=4\pi \int_{\cos(\beta_\text{max})}^{\cos(\beta_\text{min})}{\frac{dx}{(1-x)^2}(1+6x^2+x^4)\mathcal{P}_\ell(x)}\,,
\end{equation*}
\begin{equation*}
      \mathcal{K}_{\ell}^{\mp 4}=    2\pi(8!)^{\mp 1/2}\int_{\cos(\beta_\text{max})}^{\cos(\beta_\text{min})}{\frac{dx}{(1-x)^2}\sqrt{\frac{(\ell\pm 4)!}{(\ell\mp 4)!}}\mathcal{P}^{\mp 4}_\ell(x)\mathcal{P}^{\pm 4}_4(x)}\,.  
\end{equation*}
 Here, $\beta_\text{min}$ and $\beta_\text{max}$ represent the cut off range as defined in Sec. \ref{subsec:Boltzmann_equations}, $\mathcal{P}_l(x)$ and $\mathcal{P}^{\pm 4}_l(x)$ are the Legendre and associated Legendre polynomials, respectively, and using their properties it is possible to show that $\mathcal{K}_{\ell}^{+4}=\mathcal{K}_{\ell}^{-4}\equiv \mathcal{K}_{\ell}^{(4)}$.

For the sake of simplicity, we now neglect the scalar perturbations of the metric in the time interval $[\eta_{in},\eta_0]$ 
as in this toy model we are interested in estimating the contribution, to the intensity and polarisation angular power spectra observed today, of the collision terms alone. Moreover, we assume that the initial distribution of perturbations in the SGWB intensity $\Delta_I(\eta_{in})$ is statistically isotropic, $\Delta_I(\eta_{in},q,\textbf k, \textbf n)=\Delta_I(\eta_{in},q,\textbf k)$, i.e. it is represented by a monopole at the initial time $\eta_{in}$, such that different moments measured by the observer at $\eta_0$ comes only from projection effects. 

We introduce the \textbf{two-point correlation function} of the initial intensity anisotropy in Fourier space as (e.g.~\cite{Bartolo2020})
\begin{equation}
\label{eq:PS}
    \langle \Delta_I(\eta_{in},q,\textbf k)\Delta_I^*(\eta_{in},q,\textbf k')\rangle=\frac{2\pi^2}{k^3}P_{I,in}(q,k)(2\pi)^3 \delta^{(3)}(\textbf k -\textbf k')\,,
\end{equation}
with $P_{I,in}(q,k)$ the initial power spectrum, carrying out information about the physical processes generating the SGWB.

The intensity angular power spectrum observed today at the spacetime point $(\eta_0,\textbf x_0)$ is then obtained by decomposing the RHS of Eqs.~\eqref{eq:solIntensity}-\eqref{eq:solpolarized} in terms of scalar and spin-weighted spherical harmonics with $s=4$, respectively:
\begin{equation}
\begin{split}
  &\langle \Delta_{I,\ell m}(\eta_{0},q,\textbf x_0)\Delta_{I,\ell'm'}^*(\eta_{0},q,\textbf x_0)\rangle=\\
  &\int{\frac{d^3 k}{(2\pi)^3}}\int{\frac{d^3 k'}{(2\pi)^3}e^{i\textbf x_0(\textbf k - \textbf k')}}\int{d\Omega_n Y^*_{\ell m}(\textbf{n}) \Delta_{I}^\text{RHS}(\textbf n)}\int{d\Omega_n' \left[\Delta_{I}^\text{RHS}(\textbf n')\right]^*Y_{\ell' m'}(\textbf n')}\,,
  \end{split}
\end{equation}
\begin{equation}
\begin{split}
  &\langle \Delta_{Q \pm iU,\ell m}(\eta_{0},q,\textbf x_0)\Delta_{Q \pm iU,\ell'm'}^*(\eta_{0},q,\textbf x_0)\rangle=\\
  &\int{\frac{d^3 k}{(2\pi)^3}}\int{\frac{d^3 k'}{(2\pi)^3}e^{i\textbf x_0(\textbf k - \textbf k')}}\int{d\Omega_n {}_{\pm 4}Y^*_{\ell m}(\textbf{n}) \Delta_{Q \pm iU}^\text{RHS}(\textbf n)}\int{d\Omega_n' \left[\Delta_{Q \pm iU}^\text{RHS}(\textbf n')\right]^*{}_{\pm 4}Y_{\ell' m'}(\textbf n')}\,,  
  \end{split}
\end{equation}
where we have generally indicated with $\Delta_X^\text{RHS}$, with $X=I,Q \pm iU$, the RHS of Eqs.~\eqref{eq:solIntensity}-\eqref{eq:solpolarized}. After some mathematical manipulations, summarised in Appendix \ref{app:comp}, the above equations can be rearranged to give:
\begin{equation} \label{eq:PSI} 
\begin{split}
&\langle \Delta_{I,\ell m}\Delta_{I,\ell'm'}^*\rangle=
\delta_{\ell,\ell'}\delta_{m,m'}4\pi\int_0^\infty{\frac{\text{d}k}{k}P_{I,in}(k,q)\sum_{\ell_1,\ell_2}\sum_{\ell_3,\ell_4}(-i)^{\ell_1+\ell_2-\ell_3-\ell_4}}(2\ell_1+1)(2\ell_2+1)\times\\
&(2\ell_3+1)(2\ell_4+1)j_{\ell_1}[k(\eta_{in}-\eta_*)]j_{\ell_3}[k(\eta_{in}-\eta_*)]j_{\ell_2}[k(\eta_*-\eta_0)]j_{\ell_4}[k(\eta_*-\eta_0)]\times\\
&\begin{pmatrix}
  \ell_1 & \ell_2 & \ell \\
  0 & 0 & 0 
 \end{pmatrix}^2
 \begin{pmatrix}
  \ell_3 & \ell_4 & \ell \\
  0 & 0 & 0 
 \end{pmatrix}^2
\left[e^{-2\tau(\eta_{in})}+\left(\frac{\tau(\eta_*)}{1+\tau(\eta_*)}\right)^2\frac{\mathcal{K}_{\ell_1}\mathcal{K}_{\ell_3}}{\mathcal{T}^2}+\frac{2}{\mathcal{T}}\frac{\tau(\eta_*)}{1+\tau(\eta_*)}e^{-\tau(\eta_{in})}\mathcal{K}_{\ell_1}\right]\\
 &\,\,\,\,\,\,= \delta_{\ell,\ell'}\delta_{m,m'} C_{\ell,I}(\eta_{0},q)\,,
\end{split}
\end{equation}
\begin{equation} \label{eq:PSP} 
\begin{split}
&\langle \Delta_{Q \pm iU,\ell m}\Delta_{Q\pm i U,\ell'm'}^*\rangle= \delta_{\ell,\ell'}\delta_{m,m'}
\frac{512}{35}\frac{4\pi}{\mathcal{T}^2}\int_0^\infty{\frac{\text{d}k}{k}P_{I,in}(k,q)\sum_{\ell_1,\ell_2}\sum_{\ell_3,\ell_4}(-i)^{\ell_1+\ell_2-\ell_3'-\ell_4}}(2\ell_1+1)(2\ell_2+1)\times\\
&(2\ell_3+1)(2\ell_4+1)j_{\ell_1}[k(\eta_{in}-\eta_*)]j_{\ell_3}[k(\eta_{in}-\eta_*)]j_{\ell_2}[k(\eta_*-\eta_0)]j_{\ell_4}[k(\eta_*-\eta_0)]\times\\
&\begin{pmatrix}
  \ell_1 & \ell_2 & \ell \\
  0 & 0 & 0 
 \end{pmatrix}
 \begin{pmatrix}
  \ell_3 & \ell_4 & \ell \\
  0 & 0 & 0 
 \end{pmatrix}
 \begin{pmatrix}
  \ell & \ell_1 & \ell_2 \\
  \pm4 & \mp 4 & 0 
 \end{pmatrix}
 \begin{pmatrix}
  \ell & \ell_3 & \ell_4 \\
  \mp 4 & \pm 4 & 0 
 \end{pmatrix}
\left[\left(\frac{\tau(\eta_*)}{1+\tau(\eta_*)}\right)^2\mathcal\mathcal{K}_{\ell_1}^{\mp 4}\mathcal{K}_{\ell_3}^{\mp 4}\right]\\
 &\,\,\,\,\,\,= \delta_{\ell,\ell'}\delta_{m,m'} C_{\ell,Q\pm i U}(\eta_{0},q)\,.
\end{split}
\end{equation}
Note the presence of the Wigner 3j symbols:
\begin{equation*}
   \begin{pmatrix}
  j_1 & j_2 & j_3\\
  m_1 & m_2 & m_3 
 \end{pmatrix},\,  
\end{equation*}
which are connected to the Clebsch-Gordan coefficients and satisfy the triangular relation $|j_1-j_2|\le j_3 \le j_1+j_2$, together with $m_1+m_2+m_3=0$. 

The expression of $C_{\ell,I}$ and $C_{\ell,Q\pm i U}$ are quite cumbersome, even in the simplified scenario considered. In both Eqs.~\eqref{eq:PSI}-\eqref{eq:PSP}, the summed indices $\ell_1, \ell_3$ come from the incoming intensity monopole at $\eta_{in}$ which is scattered at $\eta_*$ and projected to ($\eta_0$,$\textbf x_0$), while $\ell_2, \ell_4$ are related to the outgoing radiation freely propagating from $\eta_*$ and observed at $\eta_0$. The sums are further regulated by the selection rules of the Wigner 3j symbols.

The terms between square brackets which multiply the integral over $k$ encapsulate the effect of collisions.
More in details, Eq.~\eqref{eq:PSI} can be seen as the sum of three contributions, $C_{\ell,I}=C^\text{in}_{\ell,I}+C^\text{S}_{\ell,I}+2C^\text{in,S}_{\ell,I}$, related to the three terms in square brackets.
The first term, multiplied by the exponential $\exp[-2\tau(\eta_{in})]$, refers to the initial radiation propagating without being scattered, the second term is the contribution to the intensity power spectrum from SGWB scattering at $\eta_*$, while the last term represents the cross correlation between the two.  
Considering a single multipole at a time 
at $\eta_*$ 
, $\ell_1=\ell_3= L$, it is easy to evaluate the fractional overall contribution from GW Compton scattering with respect to the unscattered radiation in the angular power spectrum:
\begin{equation}\label{eq:deltaI}
    \delta C_\text{coll,I}^{(L)}=   \left(\frac{C_{I}-C^\text{in}_{I}}{C^\text{in}_{I}}\right)^{(L)}=\left[\left(\frac{\tau(\eta_*)}{1+\tau(\eta_*)}\right)^2\frac{\mathcal{K}_{L}^2}{\mathcal{T}^2}+\frac{2}{\mathcal{T}}\frac{\tau(\eta_*)}{1+\tau(\eta_*)}e^{-\tau(\eta_{in})}\mathcal{K}_{L}\right]e^{2\tau(\eta_{in})}\,.
\end{equation}
We remark here that all the terms in Eq.~\eqref{eq:deltaI} implicitly depend on the GW frequency (wavelength) through the Rutherford-divergent cross section $\sigma(\lambda,M)$, as already pointed out in the previous sections.

\begin{figure*}
    \centering
    \includegraphics[width=0.4\textwidth]{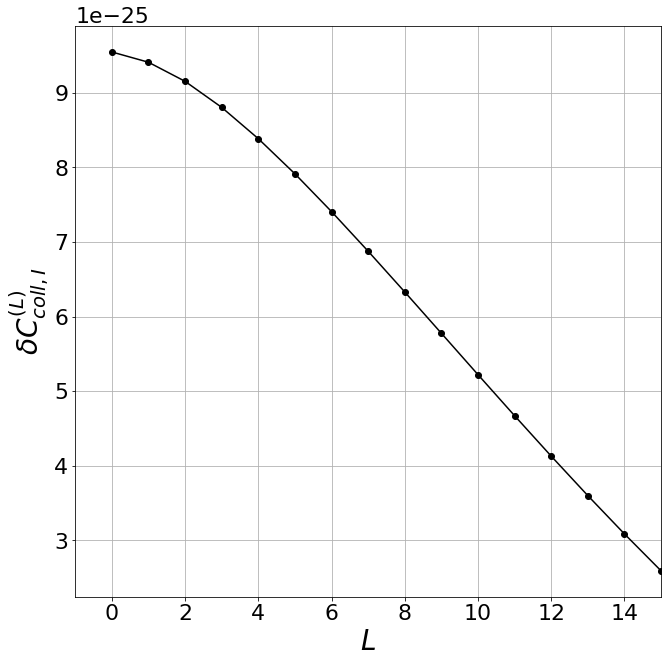}
    \includegraphics[width=0.4\textwidth]{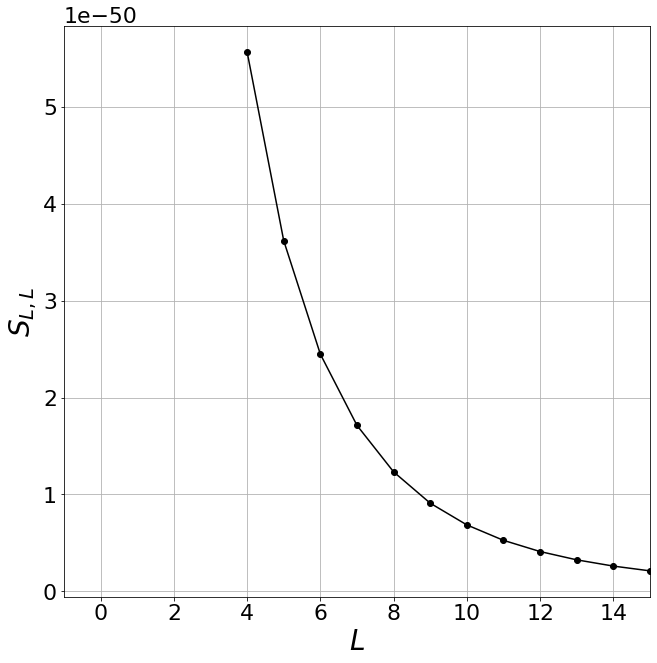}
    \caption{{Left: factional contribution of the collision term to the intensity angular power spectrum, of the intensity perturbations of a SGWB, as a function of the projected multipoles moments of the incoming radiation. Right: collision factor $S_{L,L}$ in the polarisation angular power spectrum as a function of the projected multipoles moment of the incoming radiation. The incident GWs are assumed to be an unpolarised monopole at the initial time $\eta_{in}$}.}
    \label{fig:Cli}
\end{figure*}

In the left panel of Fig.~\ref{fig:Cli}, we show $\delta C_\text{coll,I}^{(L)}$ as a function of the first 15 multipoles, for the case of a uniform distribution of stellar-mass black holes at $z_*=1.0$ with $M=10\, M_\odot$ and $\lambda_\text{obs}=10^{-10}\,\text{pc}$. As expected, the contribution of the GW Compton scattering to the intensity angular power spectrum is extremely small. Moreover it drops rapidly when going to higher multipoles. We also observe that $\delta C_\text{coll,I}^{(L)}$ is dominated by the second term which arises from cross correlation between the incoming and scattered radiation. 

The structure of Eq.~\eqref{eq:PSP} is very similar to Eq.~\eqref{eq:PSI}, except for the presence of the three j symbols with $m_1,m_2=\pm 4$. This is related to the spin 4-nature of $Q\pm i U$ and makes the term identically zero for $\ell_i<4$, $i=1..4$, as expected. The term in the square brackets of Eq.~\eqref{eq:PSP}:  
\begin{equation}
S_{\ell_1,\ell_3}=\frac{512}{35}\frac{1}{\mathcal{T}^2}\times \left[\left(\frac{\tau(\eta_*)}{1+\tau(\eta_*)}\right)^2\mathcal{K}_{\ell_1}^{(4)}\mathcal{K}_{\ell_3}^{(4)}\right]
\end{equation}
represents the order of magnitude of the polarisation generated by GW Compton scattering by massive objects. We can proceed as above and evaluate the contribution to the angular power spectrum for a single multipole, $L$, at a time. In the right panel of Fig.~\ref{fig:Cli} we show the value of $S_{L,L}$ of $Q\pm i U$ for $L=4,... 15$, confirming that the amount of polarisation generated during a single scattering process is very small for wavelengths in the LIGO-VIRGO band, as already found in~\cite{Cusin} using a different approach.

However, we stress here that the calculations presented in this section neglect scalar metric perturbations and their time derivative which affect the SGWB free-streaming from $\eta_*$ to $\eta_0$. On the one hand, as pointed out in the literature (see e.g. ~\cite{Bartolo2019}), metric perturbations produce intensity anisotropies of an even isotropic SGWB component. In turn, such intensity anisoptropies, changing in amplitude due to propagation on a perturbed background, enter the RHS of the coupled Boltzmann equations for intensity and polarisation, causing, at least via the change of $P_{I,in}(q,k)$ during GW free-streaming among different collisions (but also with the generation of the $\mathcal L^{(4)\, \pm}_m$ terms), the change of the RHS in Eqs.~\eqref{eq:PSI}-\eqref{eq:PSP}, and therefore of the observed intensity and polarisation power spectra. 
On the other hand, despite we have worked with several simplifications, the formalism we developed here is completely general and independent of the initial conditions $P_{in}(q,k)$. Therefore our computation could be extended to the case of an angular-dependent initial intensity perturbation and to polarised incoming radiation. In principle, via our approach, one could also estimate the contribution of multiple scattering episodes by considering a series of independent collisions happening at different fixed redshift $z_i$ and assuming that the SGWB evolves freely between $z_i$ and $z_{i+1}$.

\color{black}

\section{Conclusions}
In this paper we have provided, for the first time to date, a system of \textit{coupled} Boltzmann equations describing the intensity and polarisation Stokes parameters of the astrophysical SGWB, including collision terms which account for gravitational Compton scattering off of massive objects. 
In analogy with Thomson scattering in the electromagnetic case, 
we have computed the full scattering matrix, Eq.~\eqref{eq:GravitonScatteringMatrix}, appearing in the collision term of the graviton Boltzmann equations in the low-energy limit of the graviton-scalar cross-section. The final set of equations, Eqs.~\eqref{eq:IntensityEquation}-\eqref{eq:CircularPolarizationEquation}, resembles that for the CMB Stokes parameters~\cite{Kosowsky95,Kamionkowski,Durrer2008};  however, the different spin nature of the radiation and the physics involved in the scattering process determine crucial differences between photon and graviton stochastic backgrounds.

In particular, while photon Thomson scattering affects only anisotropies in the intensity of the incoming radiation with $\ell=0,2$, due to orthogonality of spherical harmonics, in the case of gravitational Compton scattering with massive objects the Rutherford's type term, $\sin^{-4}\beta$, makes all multipoles (with $m=0$) to be scattered out, therefore producing outgoing anisotropies of any order $\ell$ if they are present in the intensity of the incident GW field.

Linear polarisation can be generated from unpolarised anisotropic radiation only with $m=\pm 4$, which requires at least an hexadecapole anisotropy ($\ell\ge 4$) in the incoming intensity; this is again a consequence of the fact that $Q\pm iU$ is a spin-4 quantity in the case of gravitons.

{We have further provided an analytic solution for the angular power spectrum of intensity and polarisation anisotropies when the scattering targets are located in a small redshift range, under the assumption of unpolarised incoming monopole radiation and neglecting free-streaming effects due to metric scalar perturbations We explicitly computed the order of magnitude of the contribution of SGWB Compton scattering by massive objects to the intensity and polarisation power spectra as a function of the projected multipoles moments of the incoming radiation.}

Our analysis confirms that the contribution of the gravitational Compton scattering to SGWB anisotropies is extremely small for collisions with massive compact objects (BH and SMBH) in the frequency range of current and upcoming surveys, as already found in previous studies with different methods~\cite{Cusin}. This statement can be extended to other massive scatters as main sequence stars.
{However, in our simplified estimation of such contribution, we completely neglected effects sourced by the generation of further intensity and polarisation anisotropies during GW propagation on a perturbed FLRW background, which in turn enter the collision term in the RHS of the Boltzmann equations.}

{In conclusion, it is important to stress that the focus of this work relies on the set of coupled Boltzmann equations for SGWB intensity and polarisation Stokes parameters, which we present here for the first time in the literature. Beyond the simplified model described above, a full numerical integration of the coupled Eqs.~\eqref{eq:IntensityEquation}--\eqref{eq:CircularPolarizationEquation} would provide a more accurate estimate of the total amount of anisotropies generated by multiple scattering processes off of massive objects, as well as the interplay between polarisation and intensity (including a polarised incoming background), during the SGWB propagation across the LSS of the universe. This will be investigated in future works.}

\acknowledgments
CB and CC warmly thank Andrew Jaffe for very useful and interesting discussions since the initial concepts in 2006. LP ackwnoledges GBR and VB for useful comments and discussions.
CB acknowledges support from the InDark INFN initiative. LP acknowledges support from the Czech Academy of Sciences under the grant number LQ100102101.
LP and MC are partially supported by a 2019, 2020, 2021 and 2022 ``Research and Education'' grant from Fondazione CRT. The OAVdA is managed by the Fondazione Cl\'ement Fillietroz-ONLUS, which is supported by the Regional Government of the Aosta Valley, the Town Municipality of Nus and the ``Unit\'e des Communes vald\^otaines Mont-\'Emilius''.

\bibliographystyle{JHEP}
\bibliography{ref}

\appendix
\section{\boldmath Spin-weighted spherical harmonics for $s=\pm 4$}
\label{app:spinw}
To expand a spin-s component of a tensor field on the sphere, one employs the \textit{spin weighted spherical harmonics} (SWSH), first introduced by \cite{Newman}, which are spin-s component of tensor field on the 2-sphere. In the standard spherical basis they are given in terms of irreducible representations of the rotation group by the Goldberg formula:
\begin{equation}\label{eq:Goldberg}
\begin{split}
&{}_sY_{lm}(\theta, \phi)   = (-1)^m \sqrt{\frac{(\ell+m)!(\ell-m)!(2\ell+1)}{4\pi(\ell+s)!(\ell-s)!}}\sin^{2\ell}\Big(\frac{\theta}{2}\Big)\times \\
&\sum \binom{\ell-s}{r}\binom{\ell+s}{r+s-m}(-1)^{\ell-r-s} e^{im\phi}\cot^{2r+s-m}\Big(\frac{\theta}{2}\Big),
\end{split}
\end{equation}

where the sum runs on all values of $r$ for which the binomials are non vanishing, which means
$\text{max}(0, m-s) \leq r \leq \text{min}(\ell-s, \ell+m) $.
These functions are defined only for $\ell \geq |s|$ and $|m|\leq \ell$ as ordinary spherical harmonics. For each spin they form a comlpete orthogonal set, so that
\begin{equation}
\begin{split}
& \int \text{d}\Omega_\textbf{n} {}_sY_{\ell m}(\textbf{n}) {}_sY^*_{\ell^\prime m^\prime}(\textbf{n}) = \delta_{\ell \ell^\prime}\delta_{mm^\prime} \\
&\sum_{\ell m} {}_sY^*_{\ell m}(\textbf{n}^\prime){}_sY_{\ell m}(\textbf{n}) = \delta(\cos\theta - \cos\theta^\prime)\delta(\varphi - \varphi^\prime)\,.
\end{split}
\end{equation}

As an important property, they satisfy the \textit{generalized addition theorem} (e.g.~\cite{Hu}), which states that
\begin{equation}\label{eq:addition}
\sqrt \frac{4\pi}{2\ell+1}\sum_{m^\prime}  {}_sY_{\ell m^\prime}(\textbf{n}_2){}_{-m}Y^*_{\ell m^\prime}(\textbf{n}_1) ={}_sY_{\ell m}(\beta, \alpha) e ^{is\gamma} 
\end{equation}

where the triplet $(\alpha, \beta, \gamma)$ indentifies the Euler angles of the rotation that brings the vector $\textbf{n}_2$ on $\textbf{n}_1$.

Last, two useful properties that can be derived from the Goldberg formula are
\begin{equation}
\begin{split}
& {}_sY_{\ell m}(\pi - \theta, \phi + \pi) = (-1)^{\ell} {}_{-s}Y_{\ell m}(\theta, \phi)  \\
&{}_sY^*_{\ell m} = (-1)^{s+m}{}_{-s}Y_{\ell-m}\,.
\end{split}
\end{equation}

Here we list the explicit expressions of spin-4 spherical harmonics, calculated by using the Goldberg formula \eqref{eq:Goldberg}.
\begin{equation}
\begin {split}
& {}_{\pm 4}Y_{40}(\theta, \phi) = \frac{3}{16}\sqrt\frac{35}{2\pi} \sin^4\theta \\
& {}_{4}Y_{4\pm 4}(\theta, \phi) = \frac{1}{16}\sqrt\frac{9}{4\pi} (1 \mp \cos\theta)^4 \, e^{\pm 4i\phi} \\
& {}_{-4}Y_{4\pm 4}(\theta, \phi) = \frac{1}{16}\sqrt\frac{9}{4\pi} (1 \pm \cos\theta)^4 \, e^{\pm 4i\phi} 
\end{split}
\end{equation}

\section{Graviton-scalar scattering matrix for massive objects}
\label{app:matrix}

Using the definitions of the gravitational Stokes parameters in Sec.~\ref{sec:Stokes}, the differential cross-section in Sec.~\ref{sec:compton} and Eqs.~\eqref{eq:ScatteredAmplitudes}, we can express the scattered Stokes vector $\vec{\bm{\mathcal S}}$ in terms of the incident one as 
\begin{equation}
    \vec{\bm{\mathcal S}}=\mathcal M_{\rm SPG}\vec{\bm{\mathcal S}}^\prime,
\end{equation}
where the \textit{scattering matrix} reads as 
\begin{equation}
 \mathcal M_{\rm SPG} = \frac{1}{2 \sin^4\beta/2} \left(\mathcal A_\text{SPG}+\mathcal D_\text{SPG}\right),
\end{equation}

\begin{equation}
\begin{split}
&A_\text{SPG}=\begin{pmatrix}
1 + 6\cos^2 \beta + \cos^4\beta & \frac{1}{2}\sin^4\beta & \frac{1}{2}\sin^4\beta & 0  \\
\sin^4\beta & \frac{1}{2}(1+\cos\beta)^4 & \frac{1}{2}(1-\cos\beta)^4 & 0\\
\sin^4\beta & \frac{1}{2}(1-\cos\beta)^4 & \frac{1}{2}(1+\cos\beta)^4 & 0 \\
0 & 0 & 0 & 0
\end{pmatrix},\\
&\mathcal D_\text{SPG}=4\,\text{diag}(0,0,0,\cos\beta+\cos^3\beta)\,.
\end{split}
\end{equation}
The label SPG on the matrices refers to the fact that its expression has been found in the scattering plane geometry. 
To go from the scattering plane basis to the spherical basis in real space,
one must perform rotations in the plane orthogonal to the propagating directions, both for the incoming and the outgoing wave. In general, these are given by two different angles $\gamma$ and $\gamma^\prime$ (see e.g.~\cite{Durrer2008}), and using the transformation properties of the Stokes parameters we have
\begin{align}
\label{eq:matrixRotated}
\mathcal M = R(-\gamma) \mathcal M_{\rm SPG}, R(\gamma^\prime)= \frac{1}{2 \sin^4\beta/2} \left(\mathcal A+\mathcal D\right),
\end{align}
{where each rotation matrix $R(\alpha)$ brings a factor $\exp[\pm 4i\alpha]$ to $Q \pm iU$, due to their spin-4 nature.} 
\begin{equation}\label{eq:scatter_spherical}
\resizebox{0.9\textwidth}{!}{$
\begin{split}
&\mathcal A =\frac{8}{5}\sqrt\frac{4\pi}{9} \times\begin{pmatrix}
 6 Y_{00}(\beta, \gamma^\prime) + 12\frac{\sqrt 5}{7} Y_{20}(\beta, \gamma^\prime)+ \frac{1}{7} Y_{40}(\beta, \gamma^\prime)& 
 \frac{1}{2}\sqrt\frac{10}{7}Y_{4-4}(\beta, \gamma^\prime) & \frac{1}{2}\sqrt\frac{10}{7} Y_{44}(\beta, \gamma^\prime) & 0\\
   \sqrt\frac{10}{7}\,{}_{4}Y_{40}(\beta, \gamma^\prime) e^{-i4\gamma} &  5\,
   {}_{4}Y_{4-4}(\beta, \gamma^\prime) e^{-i4\gamma} &  5\,{}_{4}Y_{44}(\beta, \gamma^\prime) e^{-i4\gamma} & 0\\
   \sqrt\frac{10}{7}\,{}_{-4}Y_{40}(\beta, \gamma^\prime) e^{i4\gamma} &  5\,
   {}_{-4}Y_{4-4}(\beta, \gamma^\prime) e^{i4\gamma} &  5\,{}_{-4}Y_{44}(\beta, \gamma^\prime) e^{i4\gamma} & 0\\
   0 & 0 & 0 & 0\\
\end{pmatrix},\\
&\mathcal{D}=\frac{16}{5}\sqrt{\pi}\times \text{diag}\left[0,0,0,\left(\frac{10}{\sqrt{3}}Y_{10}(\beta,\gamma^\prime)+\frac{1}{\sqrt{7}}Y_{30}(\beta,\gamma^\prime)\right)\right]. \\
\end{split}
$}
\end{equation}
Eq.~\eqref{eq:scatter_spherical} can be further expressed in terms of the incoming and outgoing directions ${\bm n}^\prime$ and  ${\bm n}$ in spherical basis
by means of the generalised addition theorem for spin-weighted spherical harmonics of Eq.~\eqref{eq:addition}.
This way, we obtain the form of the scattering matrix presented in Eq.~\eqref{eq:GravitonScatteringMatrix}.

\section{Derivation of the angular power spectrum for intensity and polarisation anisotropies}
\label{app:comp}
We start by computing the coefficients $\Delta_{I,\ell m}$ of the right-hand side of Eq.~\eqref{eq:Ineta} as
\begin{equation*}
\Delta_{I,\ell m}(\eta_{0},q,\textbf x_0)=
  \int{\frac{\text{d}^3 k}{(2\pi)^3}e^{i\textbf x_0\textbf k}}\int{\text{d}\Omega_n Y^*_{\ell m}(\textbf{n}) \Delta_{I}^\text{RHS}(\eta_{0},q,\textbf k, \textbf n)}\,.
\end{equation*}
Under the assumption that the initial distribution is isotropic, and neglecting the scalar perturbations of the metric, we obtain
\begin{equation}
\begin{split}
    \Delta_{I,\ell m}=&\int{\frac{\text{d}^3 k}{(2\pi)^3}e^{i\textbf x_0\textbf k}}\int{\text{d}\Omega_n Y^*_{\ell m}(\textbf{n})\left[\Delta_I(\eta_{in},\textbf k,q)e^{ik\mu(\eta_{in}-\eta_*)-\tau(\eta_{in})}+\frac{1}{\mathcal{T}}\frac{\tau(\eta_*)}{1+\tau(\eta_*)}\right.}\times\\
    &\times \left.\sum_{\ell_1,m_1}\int{\text{d}\Omega_n' Y^*_{\ell_1 m_1}(\textbf{n}')}{\Delta_I(\eta_{in},\textbf k,q)e^{ik\mu'(\eta_{in}-\eta_*)}Y_{\ell_1 m_1}(\textbf n)\mathcal{K}_\ell}\right]e^{ik\mu(\eta_*-\eta_0)}\,,
\end{split}
\end{equation}
where we have explicitly separated in the exponentials the free propagation of the waves from the time of generation $\eta_{in}$ to the time of scattering $\eta_*$, and from the scattering to the present time $\eta_0$.

Now, the angular dependence $\mu= \hat{\textbf k} \cdot \textbf n$ can be decomposed in terms of spherical Bessel functions $j_l(k)$, by means of the expansion formula for plane waves:
\begin{equation*}
    e^{i \textbf k \cdot \textbf n}=4\pi\sum_{\ell,m}(-i)^{\ell}j_\ell(k)Y^{*}_{\ell,m}(\hat{\textbf k})Y_{\ell,m}(\textbf n)\,,
\end{equation*}
such that, using the orthogonality properties of spherical harmonics:
\begin{equation}
    \begin{split}
    \Delta_{I,\ell m}=&\int{\frac{\text{d}^3 k}{(2\pi)^3}e^{i\textbf x_0\textbf k}}(4\pi)^2\int{\text{d}\Omega_n Y^*_{\ell m}(\textbf{n})\Delta_I(\eta_{in},\textbf k,q)}\times\\
    & \times \sum_{\ell_1,m_1}(-i)^{\ell_1}j_{\ell_1}[k(\eta_{in}-\eta_*)]Y^*_{\ell_1,m_1}(\hat{\textbf k})Y_{\ell_1,m_1}(\textbf n)\left[e^{-\tau(\eta_{in})}+\frac{\mathcal{K}_{\ell_1}}{\mathcal{T}}\frac{\tau(\eta_*)}{1+\tau(\eta_*)}\right]\times\\
    &\times\sum_{\ell_2,m_2}(-i)^{\ell_2}j_{\ell_2}[k(\eta_*-\eta_0)]Y^*_{\ell_2,m_2}(\hat{\textbf k})Y_{\ell_2,m_2}(\textbf n)\,.
\end{split}
\end{equation}
Note that the integral over the solid angle $\text{d}\Omega_n$ involves three spherical harmonics; this is often referred as the \emph{Gaunt coefficient}, $\mathcal{G}^{\ell,\ell_1,\ell_2}_{m,m_1,m_2}$, and can be expressed in terms of the Wigner 3j-symbols:
\begin{equation}
\begin{split}
    \mathcal{G}^{\ell_1,\ell_2,\ell_3}_{m_1,m_1,m_2}=&\int Y_{\ell_1,m_1}(\textbf n) \
 Y_{\ell_2,m_2}(\textbf n) \ Y_{\ell_3,m_3}(\textbf n) \ \text{d}\Omega_n=
 \\
=&\sqrt{\frac{(2\ell_1+1)(2\ell_2+1)(2\ell_3+1)}{4\pi}}
 \times \begin{pmatrix} \ell_1 & \ell_2 & \ell_3 \\ 0 & 0 & 0 \end{pmatrix}
 \begin{pmatrix} \ell_1 & \ell_2 & \ell_3 \\ m_1 & m_2 & m_3 \end{pmatrix}\end{split}\,.
\end{equation}
Taking into account the relation between spherical harmonics and their complex conjugate, we obtain the final expression for the angular coefficients:
\begin{equation}
 \begin{split}
    \Delta_{I,\ell m}=&\int{\frac{\text{d}^3 k}{(2\pi)^3}e^{i\textbf x_0\textbf k}}(4\pi)^2\Delta_I(\eta_{in},\textbf k,q)\sum_{\ell_1,m_1}\sum_{\ell_2,m_2}(-i)^{\ell_1+\ell_2}j_{\ell_1}[k(\eta_{in}-\eta_*)]j_{\ell_2}[k(\eta_*-\eta_0)]\times\\
    & \times Y^*_{\ell_1,m_1}(\hat{\textbf k})Y^*_{\ell_2,m_2}(\hat{\textbf k})\left[e^{-\tau(\eta_{in})}+\frac{\mathcal{K}_{\ell_1}}{\mathcal{T}}\frac{\tau(\eta_*)}{1+\tau(\eta_*)}\right]\sqrt{\frac{(2\ell+1)(2\ell_1+1)(2\ell_2+1)}{4\pi}}
 \times \\
 &(-1)^{m}\begin{pmatrix} \ell & \ell_1 & \ell_2 \\ 0 & 0 & 0 \end{pmatrix}
 \begin{pmatrix}\ell & \ell_1 & \ell_2 \\ -m & m_1 & m_2 \end{pmatrix}\,.
\end{split}
\end{equation}
We can now insert the above equation in the two point correlation function to compute the angular power spectrum as
\begin{equation}
\label{eq:nastyI}
\begin{split}
  &\langle \Delta_{I,\ell m}(\eta_{0},q,\textbf x_0)\Delta_{I,\ell'm'}^*(\eta_{0},q,\textbf x_0)\rangle=\\
  &(4\pi)^4\int{\frac{d^3 k}{(2\pi)^3}}\int{\frac{d^3 k'}{(2\pi)^3}e^{i\textbf x_0(\textbf k - \textbf k')}}\langle \Delta_I(\eta_{in},\textbf k,q)\Delta_I^*(\eta_{in},\textbf k',q)\rangle\sum_{\ell_1,m_1}\sum_{\ell_2,m_2}\sum_{\ell_3,m_3}\sum_{\ell_4,m_4} \\
  & (-i)^{\ell_1+\ell_2-\ell_3-\ell_4}j_{\ell_1}[k(\eta_{in}-\eta_*)]j_{\ell_2}[k(\eta_*-\eta_0)]j_{\ell_3}[k'(\eta_{in}-\eta_*)]j_{\ell_4}[k'(\eta_*-\eta_0)]\times\\
    & \times (-1)^{m+m'}Y^*_{\ell_1,m_1}(\hat{\textbf k})Y^*_{\ell_2,m_2}(\hat{\textbf k})Y_{\ell_3,m_3}(\hat{\textbf k}')Y_{\ell_4,m_4}(\hat{\textbf k}')\sqrt{\frac{(2\ell+1)(2\ell_1+1)(2\ell_2+1)}{4\pi}}\times\\
    &\times\sqrt{\frac{(2\ell'+1)(2\ell_3+1)(2\ell_4+1)}{4\pi}}\begin{pmatrix} \ell & \ell_1 & \ell_2 \\ 0 & 0 & 0 \end{pmatrix}
 \begin{pmatrix}\ell & \ell_1 & \ell_2 \\ -m & m_1 & m_2 \end{pmatrix}\begin{pmatrix} \ell' & \ell_3 & \ell_4 \\ 0 & 0 & 0 \end{pmatrix}
 \begin{pmatrix}\ell & \ell_1 & \ell_2 \\ -m' & m_3 & m_4 \end{pmatrix}\times\\
 &\times\left[e^{-2\tau(\eta_{in})}+\left(\frac{\tau(\eta_*)}{1+\tau(\eta_*)}\right)^2\frac{\mathcal{K}_{\ell_1}\mathcal{K}_{\ell_3}}{\mathcal{T}^2}+\frac{2}{\mathcal{T}}\frac{\tau(\eta_*)}{1+\tau(\eta_*)}e^{-\tau(\eta_{in})}\mathcal{K}_{\ell_1}\right] \,.
  \end{split}
\end{equation}
The expression above appears to be quite complicated, but it can be slightly simplified with some useful observations. First, the correlator $\langle \Delta_I(\eta_{in},\textbf k,q)\Delta_I^*(\eta_{in},\textbf k',q)\rangle$ is given by Eq.~\eqref{eq:PS}, where the Dirac delta can be used to remove one of the integrals over $\textbf k$. 
Then, one has to deal with the integral of four spherical harmonics, which can be written again in terms of Wigner 3j-symbols:
\begin{equation}
\begin{split}
   & \int Y_{\ell_1m_1}(\textbf k) \, Y_{\ell_2m_2}(\textbf k) \, Y_{\ell_3m_3}(\textbf k) \, Y_{\ell_4m_4}(\textbf k)\text{d}\Omega_k =\\
& =  (-1)^{m_1+m_2} \, \frac{\sqrt{(2\ell_1+1)(2\ell_2+1)(2\ell_3+1)(2\ell_4+1)}}{4\pi} \sum_{\ell} (2\ell+1) \, \begin{pmatrix}\ell_1&\ell_2&\ell\\0&0&0\end{pmatrix} \, \begin{pmatrix}\ell_3&\ell_4&\ell\\0&0&0\end{pmatrix}\\ &\times\begin{pmatrix}\ell_1&\ell_2&\ell\\m_1&m_2&m_3+m_4\end{pmatrix} \, \begin{pmatrix}\ell_3&\ell_4&\ell\\m_3&m_4&m_1+m_2\end{pmatrix}\,.
\end{split}
\end{equation}
Note that the dependence on $m_i$ in Eq.~\eqref{eq:nastyI} is all in the Wigner 3j-symbols. With the help of the summation properties and symmetries of Wigner 3j-symbols, it is therefore possible to use the sum over $m_1,m_2,m_3,m_4$ to obtain $\delta_{m,m'}\delta_{\ell,\ell'}$, which finally brings Eq.~\eqref{eq:nastyI} into the result of Eq.\eqref{eq:PSI}.

The derivation of the angular power spectrum for linear polarisation anisotropies is very similar to the one we have shown in the case of intensity, but with the important caveat that the spin-4 nature of $Q\pm i U$ requires a decomposition in terms of spin-weighted spherical harmonics. In this case, the expression for the $\Delta_{Q\pm iU,\ell,m}$ coefficients becomes:
\begin{equation}
    \begin{split}
    \Delta_{Q\pm iU,\ell,m}=&16\sqrt{\frac{2}{35}}\int{\frac{\text{d}^3 k}{(2\pi)^3}e^{i\textbf x_0\textbf k}}(4\pi)^2\int{\text{d}\Omega_n \,{}_{\pm 4}Y^*_{\ell m}(\textbf{n})\Delta_{I,\ell,m}(\eta_{in},\textbf k,q)}\times\\
    & \times \sum_{\ell_1,m_1}(-i)^{\ell_1}j_{\ell_1}[k(\eta_{in}-\eta_*)]\,Y^*_{\ell_1,m_1}(\hat{\textbf k}){}_{\pm 4}Y_{\ell_1,m_1}(\textbf n)\left[\frac{\mathcal{K}_{\ell_1}^{(4)}}{\mathcal{T}}\frac{\tau(\eta_*)}{1+\tau(\eta_*)}\right]\times\\
    &\times\sum_{\ell_2,m_2}(-i)^{\ell_2}j_{\ell_2}[k(\eta_*-\eta_0)]Y^*_{\ell_2,m_2}(\hat{\textbf k})Y_{\ell_2,m_2}(\textbf n)\,.
\end{split}
\end{equation}
The integral over $\text{d}\Omega_n$ now involves two spin-weighted $s=\pm 4$ and one scalar spherical harmonics; we can then use the fact that
\begin{equation}
\begin{split}
    \int_{S^2} \text{d}\Omega\,{}_{s_1} Y_{j_1 m_1}
\,{}_{s_2} Y_{j_2m_2}\, {}_{s_3} Y_{j_3m_3} &=\\
&\sqrt{\frac{\left(2j_1+1\right)\left(2j_2+1\right)\left(2j_3+1\right)}{4\pi}}
\begin{pmatrix}
  j_1 & j_2 & j_3\\
  m_1 & m_2 & m_3
\end{pmatrix}
\begin{pmatrix}
  j_1 & j_2 & j_3\\
  -s_1 & -s_2 & -s_3
\end{pmatrix}\,,
\end{split}
\end{equation}
if $s_1+s_2+s_3=0$, as it is in our case, to get the expression:
\begin{equation}\label{eq:anicoeff}
    \begin{split}
    \Delta_{Q\pm iU,\ell,m}=&16\sqrt{\frac{2}{35}}\int{\frac{\text{d}^3 k}{(2\pi)^3}e^{i\textbf x_0\textbf k}}(4\pi)^2\Delta_I(\eta_{in},\textbf k,q)\sum_{\ell_1,m_1}\sum_{\ell_2,m_2}(-i)^{\ell_1+\ell_2}(-1)^{m}\times\\
    & \times j_{\ell_1}[k(\eta_{in}-\eta_*)] j_{\ell_2}[k(\eta_*-\eta_0)] Y^*_{\ell_1,m_1}(\hat{\textbf k})Y^*_{\ell_2,m_2}(\hat{\textbf k})\left[\frac{\mathcal{K}_{\ell_1}^{(4)}}{\mathcal{T}}\frac{\tau(\eta_*)}{1+\tau(\eta_*)}\right]
 \times \\
 &\times \sqrt{\frac{(2\ell+1)(2\ell_1+1)(2\ell_2+1)}{4\pi}}\begin{pmatrix} \ell & \ell_1 & \ell_2 \\ \pm 4 & \mp 4 & 0 \end{pmatrix}
 \begin{pmatrix}\ell & \ell_1 & \ell_2 \\ -m & m_1 & m_2 \end{pmatrix}\,.
\end{split}
\end{equation}
The angular power spectrum of Eq.~\eqref{eq:PSP} can now be computed by multiplying Eq.~\eqref{eq:anicoeff} by its complex conjugate and performing the ensamble average, following the same steps discussed above for the intensity case.

\color{black}
\end{document}